\documentclass[onecolumn,draftcls,12pt]{IEEEtran}

\usepackage{amssymb}
\usepackage{amsmath}
\usepackage{cite}
\usepackage{url}
\usepackage{xcolor}
\usepackage{cite,graphicx,amsmath,amssymb}
\usepackage{fancyhdr}
\usepackage{mdwmath}
\usepackage{mdwtab}
\usepackage{caption}
\usepackage{amsthm}
\usepackage{xcolor}
\usepackage{bm}
\usepackage{amsmath}   
\usepackage{amssymb}   
\usepackage{mathrsfs}
\usepackage{setspace}
\usepackage{algorithm}
\usepackage{algorithmic}
\usepackage{pdfpages}
\usepackage{mathtools}
\usepackage{subcaption}
\usepackage{bbm}
\usepackage{comment}
\usepackage{tikz}
\usetikzlibrary{arrows.meta,positioning}

\graphicspath{{./src/img/}}
\usepackage{xcolor} 
\usepackage[colorlinks=true, linkcolor=black, urlcolor=black,citecolor=blue]{hyperref} 
\usepackage{booktabs}
\usepackage{makecell}

\newtheorem{theorem}{Theorem}

\newtheorem{lemma}{Lemma}

\newtheorem{corollary}{Corollary}

\newtheorem{proposition}{Proposition}

\newtheorem{example}{Example}

\allowdisplaybreaks
\expandafter\def\expandafter\normalsize\expandafter{%
    \normalsize%
    \setlength\abovedisplayskip{4pt}%
    \setlength\belowdisplayskip{4pt}%
    \setlength\abovedisplayshortskip{2pt}%
    \setlength\belowdisplayshortskip{2pt}%
}

\title{Task-Aware Frequency-Division Multiplexing for Over-the-Air Computing: Design and Analysis}
		\author{
        Biao Dong,
	    Bin Cao, \IEEEmembership{Member, IEEE},\\
        Guan Gui, \IEEEmembership{Fellow, IEEE},
        and Qinyu Zhang, \IEEEmembership{Senior Member, IEEE}
		\thanks{}
		\IEEEcompsocitemizethanks{
                \IEEEcompsocthanksitem 
			\IEEEcompsocthanksitem Biao Dong, Bin Cao and Qinyu Zhang are with the School of Electronic and Information Engineering, Harbin Institute of Technology (Shenzhen), Shenzhen 518055, China (e-mail: 23b952012@stu.hit.edu.cn; caobin@hit.edu.cn; zqy@hit.edu.cn).
			\IEEEcompsocthanksitem Guan Gui is with the College of Telecommunications and Information Engineering, Nanjing University of Posts and Telecommunications, Nanjing 210003, China (e-mail: guiguan@njupt.edu.cn).
	}
 }

\begin{document}

\maketitle

\begin{abstract}
In this paper, we consider a distributed \emph{integrated sensing, communication, and computation} (ISCC) system, 
where distributed devices perform target sensing and then offload locally extracted features to the \emph{access point} (AP) for collaborative downstream classification. We study how to maximize classification performance by leveraging \emph{over-the-air computation} (AirComp). For linear classification, the theoretical analysis reveals that the classification performance of the aggregated feature is governed by the class-averaged \emph{discriminant gain} (DG), which serves as a task-aware proxy for class separability. This raises the natural question of whether the task-aware design always outperforms conventional design. We answer this question by proving that, under \emph{time-division multiplexing} (TDM), task-aware aggregation exhibits a threshold-based structure similar to that of conventional AirComp. This structure motivates a task-aware \emph{frequency-division multiplexing} (FDM) mechanism that enables non-uniform resource allocation according to task importance. We extend this mechanism to scenarios with multiple receive antennas and imperfect \emph{channel state information} (CSI), where the corresponding receive-antenna scaling law and regularized transceiver structure are derived. For nonlinear \emph{deep neural network} (DNN) classification, an empirically parameterized design is proposed, which is guided by the insights obtained from linear classification. Experimental results verify that the proposed FDM mechanism achieves superior classification performance and strikes a tradeoff between inter-class separation and intra-class variance, while imperfect CSI diminishes the transmit power gain provided by multi-antenna diversity.

\end{abstract}


\section{Introduction}

\subsection{Background}
To support latency-sensitive applications such as smart factories, digital twins, and the low-altitude economy, the \emph{sixth generation} (6G) of wireless networks is expected to extend its functional scope by integrating sensing capabilities, thereby establishing the paradigm of \emph{integrated sensing and communication} (ISAC). In this fashion, massive sensory data needs to be processed in real time, which naturally involves three tightly coupled processes, namely sensing, communication, and computation, as illustrated in Fig.~\ref{1Multi_User_Task_ISCC_intro_black}. This gives rise to a new research area called \emph{integrated sensing, communication, and computation} (ISCC) \cite{wen2024survey}.

An efficient ISCC system relies on two key aspects: system-level optimization and sensing-level enhancement \cite{liu2025integrated}. The former focuses on the joint optimization of sensing, communication, and computation processes. The latter aims to improve data quality from the sensing source through joint optimization of sensing error and multi-device sensing aggregation. However, the performance improvement from both aspects is fundamentally constrained by the frequent transmission of massive sensory data, which results in a severe communication bottleneck.

\subsection{Related Works \& Motivation}
As a promising multiple-access technique, \emph{over-the-air computation} (AirComp) integrates multi-access transmission with nomographic functional computation (e.g., sum and averaging) by exploiting the signal superposition property of wireless multiple-access channels \cite{csahin2023survey,cao2020optimized,liu2020over}.
This enables AirComp to overcome the scalability bottleneck of conventional orthogonal multiple-access transmission, where radio resources are divided among users. 

\subsubsection{Basic AirComp} 
The idea of function computation over a multiple-access channel can be traced back to the pioneering work of Nazer and Gastpar~\cite{nazer2007computation}, where the asymptotic computation rate was investigated under the assumption of \emph{independent and identically distributed} (i.i.d.) sources. Their subsequent work showed that, uncoded analog transmission is optimal for recovering the underlying Gaussian source in terms of \emph{mean squared error} (MSE) \cite{gastpar2008uncoded}. This motivates an active area focusing on designing and implementing techniques for receiving a desired function of superimposed signals, including power control~\cite{cao2020optimized,liu2020over}, multi-antenna processing~\cite{zhu2018mimo}, device synchronization~\cite{abari2015airshare}, and robustness against \emph{channel state information} (CSI) errors~\cite{shao2022bayesian}. For further details, we refer the reader to~\cite{csahin2023survey} and the references therein.

\subsubsection{Integrated Sensing and AirComp} 
Building on the well-established AirComp technique, a series of works have explored the integration of sensing and AirComp. These studies can be broadly divided into two categories: \emph{sensing-centric design} and \emph{computation-centric design}. The sensing-centric design investigates the fundamental \emph{sensing-computation} tradeoff, where computation performance is measured by the AirComp error \cite{qi2020integrated,li2023integrated,fu2025predictive}, whereas sensing performance is characterized by various metrics such as the \emph{signal-to-interference-plus-noise ratio} (SINR) \cite{qi2020integrated}, the \emph{Cramér--Rao lower bound} (CRLB) \cite{li2023integrated}, and the \emph{posterior CRLB} (PCRLB) \cite{fu2025predictive}. The computation-centric design aims to aggregate distributed state values, thereby enabling the estimation \cite{chen2018over,huynh2026over} or detection \cite{feres2023over} of global consensus. Since the data transmitted by AirComp originates from the sensing process, efficient system resource management (e.g., power and bandwidth), is crucial for achieving efficient computation \cite{chen2018over,huynh2026over,feres2023over}.

\subsubsection{Integrated Sensing, AirComp and Downstream Task} From the computational perspective, intelligent computation is evolving toward an edge \emph{artificial intelligence} (AI) assisted offloading architecture \cite{wen2023task,wang2026revisiting,chen2024view}. A fundamental problem for this architecture is to enhance the \emph{downstream task performance} by maximizing task-relevant information (e.g., classification and target detection), thereby shifting the focus from faithful reconstruction to task-oriented feature transmission. In view of this shift, AirComp plays a key role by enabling the efficient aggregation of high-dimensional features for intelligent edge computation \cite{chen2024view}.

Existing studies have partially investigated how to maximize task performance by leveraging AirComp \cite{chen2023over,xie2023optimal,zhuang2023integrated,wen2023task,wang2026revisiting,chen2024view}. Specifically, \cite{wen2023task} proposed a task-oriented AirComp framework for ISCC, where the \emph{discriminant gain} (DG) between two classes is adopted as a task-aware proxy for classification performance. This scheme was further shown to retain its superior performance under the finite-blocklength feature transmission framework~\cite{wang2026revisiting}.
From an analytical perspective, \cite{chen2024view} refined the multi-access process and compared AirComp with \emph{orthogonal multiple access} (OMA). Their analysis demonstrates that AirComp substantially outperforms OMA and achieves superior asymptotic scaling behavior, particularly when the number of receive antennas at the \emph{access point} (AP) is limited \cite{chen2024view}. Despite their remarkable performance, most existing works have not fully exploited the potential of AirComp \cite{wen2023task,wang2026revisiting,chen2024view}. 
In particular, they typically employ single-carrier transmission under \emph{time-division multiplexing} (TDM), which, as will be demonstrated later, may limit the achievable task performance. Although multi-carrier transmission has been introduced for AirComp in \cite{zhuang2023integrated,chen2023over,xie2023optimal}, 
 these works either neglect task impacts \cite{chen2023over,xie2023optimal} or assume perfect CSI \cite{zhuang2023integrated}, thereby overlooking the transceiver scaling behavior under imperfect CSI with multiple receive antennas. Meanwhile, prior works largely focus on specific algorithmic designs \cite{chen2023over,xie2023optimal,zhuang2023integrated,wen2023task,wang2026revisiting,chen2024view}, a theoretical comparative analysis between task-aware design and classic AirComp design remains lacking; specifically, whether task-aware design always outperforms the classic AirComp design? where the potential performance gains of task-aware design originate? Clearly, these questions leave much room for investigating task-aware AirComp.

\subsection{Our Contributions}
Motivated by the aforementioned perspectives, this paper investigates a AirComp-enabled ISCC system with the objective of maximizing classification performance, as illustrated in Fig.~\ref{1Multi_User_Task_ISCC_intro_black}. We begin with a theoretical analysis of classical AirComp under linear classification and then extend the framework to the task-aware design. The key contributions of this work are summarized as follows. 
\begin{enumerate}
   \item  \emph{Task-Aware FDM for AirComp}: We analyze the classification performance in terms of class average DG for linear classification, whose geometric interpretation is the Mahalanobis distance between different classes. We then derive the optimal transceiver design for maximizing the DG under the TDM transmission. A comparison with the classical AirComp reveals that both designs exhibit similar threshold-based structures. Motivated by this observation, a task-aware \emph{frequency-division multiplexing} (FDM) mechanism is proposed, which enables non-uniform resource allocation according to the task importance. The analysis further reveals that the proposed AirComp design essentially strikes a tradeoff between inter-class separation and intra-class variance. 
   \item  \emph{CSI Acquisition and DNN Classification}: We extend the proposed task-aware AirComp mechanism to practical scenarios with multiple receive antennas and prove that increasing the number of antennas at the AP enables the devices to scale down their transmit power without degrading the DG. We develop an efficient CSI acquisition and feedback scheme tailored to the proposed design. We show that CSI errors introduce a regularization term into the transceiver design and weaken the transmit-power scaling law. Due to the high nonlinearity of \emph{deep neural networks} (DNNs), the task-aware FDM design for DNN classification is analytically intractable. Guided by the insights obtained from linear classification, an empirically parameterized design is thus proposed, which combines channel inversion with gradient-prior weighting.

\end{enumerate}

The remainder of this paper is organized as follows. Section~\ref{sec:system_model} presents the system model. Section \ref{sec:sensing_level_bayesian_analysis} analyzes the  classification performance from the perspective of linear classification. Section~\ref{sec:Task-Aware} analyzes the task-aware AirComp scheme under FDM. The CSI-related considerations and DNN classification are discussed in Section~\ref{sec:csi}. Section~\ref{sec:results} presents the numerical results, and Section~\ref{sec:conclusion} concludes the paper.

{\textit {Notations}}: Boldface letters denote vectors (e.g., $\boldsymbol{x}$). $\mathcal{N}(\mu, \sigma^2)$ represents the Gaussian distribution with mean $\mu$ and variance $\sigma^2$. The superscript $(\cdot)^{\mathsf T}$ and $(\cdot)^{\mathsf H}$ denote the transpose and conjugate-transpose operations, respectively. $H(\cdot\mid\cdot)$ represents the conditional entropy. $\mathbf{I}_{a}$ stands for the $a \times a$ identity matrix. $\mathsf{KL}(\cdot\parallel\cdot)$ represents \emph{Kullback-Leibler} (KL) divergence.

\section{System Model} \label{sec:system_model}
\begin{figure}[!t]    
	\centering
	{\includegraphics[width=0.6\textwidth]{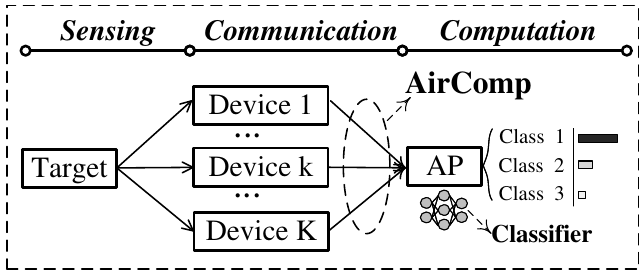}}
	\caption{Multiple devices transmit probing signals for target sensing and then offload the sensing data to the AP for computation.}
    \label{1Multi_User_Task_ISCC_intro_black}
\end{figure}

Consider a AirComp-enabled ISCC system, as illustrated in Fig.~\ref{1Multi_User_Task_ISCC_intro_black}.
A set of $K$ devices observe one target for collaborative sensing. Each device performs local feature extraction based on its observations and transmits the features to the AP over a multiple-access channel, thereby aggregating the features for collaborative classification.

\subsection{Sensing Model}
\subsubsection{Feature Distribution}
The observation of device $k$, say $\boldsymbol{r}_k=[{r}_{k,1}, {r}_{k,2}, \ldots, {r}_{k,F}]^{\mathsf T}\in \mathbb{R}^{F}$, is obtained from radar echoes, e.g., \emph{frequency-modulated continuous-wave} (FMCW) radar. Due to the bandwidth limitation, each observation $\boldsymbol{r}_k$ must be compressed before transmission, e.g., \emph{convolutional neural network} (CNN) compression. For analytical tractability, we model the compressed feature ${\boldsymbol{x}}_k = [{x}_{k,1}, \dots, {x}_{k,M}]^{\mathsf T}\in \mathbb{R}^{M}$ as a Gaussian distribution. Since the task-oriented classification involves multiple classes, a simple Gaussian distribution is insufficient to capture this complexity. We alternatively consider the \emph{Gaussian mixture} (GM) model, a superposition of $L$ class-conditional Gaussians with a uniform class prior (e.g., \cite{wen2023task,wang2026revisiting,chen2024view}). The joint distribution of the $K$ compressed features is given by
\begin{align}\label{eq:overall_Gaussian}
p(\boldsymbol{x}_1,\ldots,\boldsymbol{x}_K)=
\frac{1}{L}
\sum_{\ell=1}^{L}
p(\boldsymbol{x}_1,\ldots,\boldsymbol{x}_K \mid \ell),
\end{align}
where $\ell \in \mathcal{L} = \{1, \cdots, {L}\}$ denotes the class label. We assume that the observations are conditionally independent given the class label, i.e.,
\begin{align*}
p(\boldsymbol{x}_1,\ldots,\boldsymbol{x}_K \mid \ell)=
\prod_{k=1}^{K} p(\boldsymbol{x}_k \mid \ell).
\end{align*}
This is justified in practical collaborative sensing scenarios, where spatially separated devices observe the target from distinct views. Since the feature $\boldsymbol{x}_k$ has $M$ dimensions, it is modeled as an $M$-dimensional multivariate Gaussian distribution as
\begin{align}\label{Eq:class_distribution}  
p(\boldsymbol{x}_k\mid \ell) = \mathcal{N}\left(\boldsymbol{x}_k \mid \boldsymbol{\mu}_\ell, \boldsymbol{\Sigma}\right)
\end{align}
where $\boldsymbol{\mu}_\ell = [\mu_{\ell,1},\mu_{\ell,2},\cdots,\mu_{\ell,M}]^{\mathsf T}$ denotes the mean vector and $\boldsymbol{\Sigma}= \mathrm{diag} (\sigma_1^2,\sigma_2^2,\cdots,\sigma_{M}^2)$ denotes the diagonal covariance matrix. For the ISCC system illustrated in Fig.~\ref{1Multi_User_Task_ISCC_intro_black},
all the compressed features $\{{\boldsymbol{x}}_k\}_{k=1}^K$ are first aggregated into a single feature vector as
\begin{equation}\label{Eq:ideal_received_signal}
    \boldsymbol{y} = \frac{1}{K}\sum_{k=1}^{K} {\boldsymbol{x}}_k.
\end{equation}

Then, the aggregated feature vector $\boldsymbol{y}$ is fed into the classifier deployed at the AP for classification. Two classification models are considered, detailed below.
\begin{enumerate}
\item \emph{\underline{Linear Classification}}:
The theoretically optimal classifier follows the \emph{maximum-likelihood} (ML) rule under a uniform class prior \cite{bishop2006pattern}, i.e., 
\begin{equation}\label{eq:map_classifier}
\hat \ell = 
\arg\max_{\ell\in\mathcal L} p({\boldsymbol{y}}|\ell),
\end{equation}
where $\hat \ell$ denotes the predicted class label. Accordingly, the class-conditional likelihood is given by \cite{bishop2006pattern}
\begin{equation}\label{eq:class_conditional_likelihood}
p(\boldsymbol{y}\mid\ell)
\propto\exp\left[-\frac{K}{2}\left(\boldsymbol{y}-{\boldsymbol{\mu}_{\ell}}\right)^{\mathsf{T}}{\boldsymbol{\Sigma}}^{-1}\left(
\boldsymbol{y}-{\boldsymbol{\mu}}_{\ell}
\right)\right],
\end{equation}
which corresponds to the squared Mahalanobis distance from the vector $\boldsymbol{y}$ to the class centroid $\boldsymbol{\mu}_{\ell}$. Then, the linear classification reduces to Mahalanobis distance classification.

\item \emph{\underline{DNN Classification}}: A more realistic but analytically intractable setting is DNN classification, where a global DNN is partitioned into a device-side subnetwork and an AP-side classifier. Each device employs the device-side subnetwork to extract the feature $\{{\boldsymbol{x}}_k\}_{k=1}^K$, while the AP-side classifier performs classification based on the aggregated feature vector.

\end{enumerate}

\subsection{Communication Model}
We adopt analog AirComp to transmit the compressed features ${\boldsymbol{x}}_k$ in \eqref{Eq:class_distribution} to the AP, which exploits the signal superposition property of the wireless multiple-access channel to accomplish a specific computation \cite{cao2020optimized}. An example is the arithmetic mean, modeled as \eqref{Eq:ideal_received_signal}, 
where $\boldsymbol{y}$ denotes the ideal aggregated signal at the AP \footnote{Common computation includes averaging, maximum, and summation. Interested readers can refer to \cite{csahin2023survey} and references therein.}. 

The frequency-selective slow-fading channel is considered for all devices, where the channel remains constant over a coherence duration of $T_{\rm cd}$. To transmit the feature vector ${\boldsymbol{x}}_k = [{x}_{k,1}, \dots, {x}_{k,M}]^{\mathsf T}$ within one coherence duration, two transmission schemes are considered: TDM and FDM \cite{carleial1978interference}.
\begin{figure}[!h]
    \centering
    \includegraphics[width=0.6\textwidth]{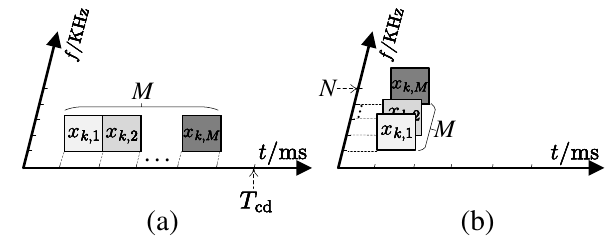}
\caption{Two transmission schemes: (a) TDM and (b) FDM}
    \label{5_Fig5_TDM_FDM}
\end{figure}
\begin{enumerate}
\item \emph{\underline{TDM Transmission}}: Each feature element is sequentially transmitted over $M$ time slots, with $M\leq T_{\rm cd}$; see Fig.~\ref{5_Fig5_TDM_FDM}(a). The received signal $\tilde{\boldsymbol{y}}^{\sf t}=[\tilde{y}_1^{\sf t},\ldots,\tilde{y}_M^{\sf t}]^{\mathsf T}$ at the AP is
\begin{equation}\label{eq:TDM_channel}
\tilde{\boldsymbol{y}}^{\sf t}
= \sum_{k=1}^{K} \boldsymbol{h}_k \odot \boldsymbol{b}_k \odot {\boldsymbol{x}}_k
+ \boldsymbol{w},
\end{equation}
where the superscript $\sf{t}$ denotes TDM, $\boldsymbol{h}_k=[h_{k,1},\ldots,h_{k,M}]^{\mathsf T}$ denotes the channel vector over $M$ slots at device $k$, $\boldsymbol{b}_k=[b_{k,1},\ldots,b_{k,M}]^{\mathsf T}$ denotes the corresponding transmit vector, $\odot$ denotes the element-wise multiplication, and $\boldsymbol{w}\sim\mathcal{CN}(\boldsymbol{0},\sigma_w^2\mathbf{I}_M)$ denotes the AWGN vector. Since $M \leq T_{\rm cd}$, the channel remains constant over the $M$ slots, i.e., $h_{k,m}=h_k$, $\forall m$. 

\item \emph{\underline{FDM Transmission}}: The feature elements are transmitted in parallel over $N$ subcarriers, with $M\leq N$; see Fig.~\ref{5_Fig5_TDM_FDM}(b). The received signal $\tilde{\boldsymbol{y}}^{\sf f}=[\tilde{y}_1^{\sf f},\ldots,\tilde{y}_M^{\sf f}]^{\mathsf T}$ at the AP is given by
\begin{equation}\label{eq:FDM_channel}
\tilde{\boldsymbol{y}}^{\sf f}
= \sum_{k=1}^{K} \boldsymbol{h}_k \odot \boldsymbol{b}_k \odot {\boldsymbol{x}}_k
+ \boldsymbol{w},
\end{equation}
With slight abuse of notation,  $\boldsymbol{h}_k$,  $\boldsymbol{b}_k$ and $\boldsymbol{w}$ in \eqref{eq:FDM_channel} denote the frequency-domain channel, transmit vector and AWGN over the $M$ subcarriers, respectively. The superscript $\sf{f}$ corresponds to FDM.
\end{enumerate}

After receiving the feature vector $\tilde{\boldsymbol{y}}^{\sf t}$ or $\tilde{\boldsymbol{y}}^{\sf f}$, the AP performs receiver aggregation. Taking the FDM transmission $\tilde{\boldsymbol{y}}^{\sf f}$ as an example, the AP applies linear aggregation with the receive vector $\boldsymbol{a}=[a_1,\ldots,a_M]^{\mathsf T}$, yielding the noisy version of the desired aggregated signal in \eqref{Eq:ideal_received_signal} as
\begin{equation}\label{Eq:aggregation_received_signal}
\hat{\boldsymbol{y}}
\!=\! \frac{1}{K}\boldsymbol{a} \odot \tilde{\boldsymbol{y}}^{\sf f}
\!=\! \frac{1}{K}\!\!\left(\!\!\boldsymbol{a} \!\odot \!\!\sum_{k=1}^{K}
\boldsymbol{h}_k \odot \boldsymbol{b}_k \odot {\boldsymbol{x}}_k
+ \boldsymbol{a} \odot \boldsymbol{w}\right).
\end{equation}
The aggregated feature vector $\hat{\boldsymbol{y}}=[\hat{y}_1,\ldots,\hat{y}_{M}]^{\mathsf T}$ is then fed into the classification model. The TDM case is analogous, where aggregation is performed across the $M$ time slots.

With the aggregated observation $\hat{\boldsymbol{y}}$ available, the AP-side classifier determines the predicted class label $\hat \ell$.
The entire process from sensing to classification forms the following Markov chain
\begin{equation*}
    \ell \rightarrow {\boldsymbol r}_k \rightarrow
	{\boldsymbol{x}}_k\rightarrow
	\boldsymbol{\tilde{y}}\rightarrow \hat{\boldsymbol{y}}\rightarrow \hat \ell.
\end{equation*}
Based on this Markov chain, we observe that, with a fixed classification model, the classification performance is jointly determined by two factors: (i) Sensing quality $\ell \rightarrow\boldsymbol{r}_k \rightarrow {\boldsymbol{x}}_k$, and (ii) Communication error ${\boldsymbol{x}}_k \rightarrow \boldsymbol{\tilde{y}} \rightarrow \hat{\boldsymbol{y}}$. In the following sections, we first theoretically analyze the classification performance for linear classification. The resulting analytical insights are then used to guide the design of nonlinear DNN classification.

\section{Aggregation Analysis}\label{sec:sensing_level_bayesian_analysis}
In this section, we analyze the aggregation performance in terms of linear classification. We first consider ideal noise-free aggregation in the absence of channel fading and noise, and then investigate noisy aggregation under TDM transmission.

\subsection{Noise-Free Aggregation Analysis}
We analyze the aggregation performance under a noise-free scenario [see \eqref{Eq:ideal_received_signal}]. The fundamental limit of the achievable aggregation performance is characterized by the conditional entropy \cite{chen2024view}, i.e., $H(\ell \mid \boldsymbol y)$; see the following lemma. 
\begin{lemma}[Aggregation Performance \cite{chen2024view}]
\label{lemma:GM_aggregation_gain}
The conditional entropy of the class label given the aggregated feature $\boldsymbol y$ can be approximated by
\begin{align*}
H\big(\ell \mid \boldsymbol y\big) \approx \log \left[1 + (L-1)\exp \left(-\frac{1}{2}
\bar{\mathcal{G}}^{\boldsymbol y}\right)\right]+C_a, 
\end{align*}
where $C_a$ denotes the second-order remainder associated with the first-order Taylor approximation and $\bar{\mathcal{G}}^{\boldsymbol y}$ denotes the class average DG, defined as
\begin{equation}\label{eq:average_DG} \bar{\mathcal{G}}^{\boldsymbol y} \triangleq \frac{2}{L(L-1)} \sum_{\ell=1}^{L} \sum_{\ell'=\ell+1}^{L} \mathcal{G}^{\boldsymbol y}_{(\ell,\ell')}. \end{equation}
Here, $\mathcal{G}^{\boldsymbol{y}}_{({\ell,\ell^{\prime}})}$ denotes the DG between classes $\ell$ and $\ell'$, defined via
the symmetric KL divergence as
\begin{align}\label{eq:symmetric_KL}
\mathcal{G}^{\boldsymbol y}_{({\ell,\ell^{\prime}})}&=\mathsf{KL}\!\left(p(\boldsymbol y\mid\ell)\|p(\boldsymbol y\mid\ell^\prime)\right)
+
\mathsf{KL}\!\left(p(\boldsymbol y\mid\ell^\prime)\|p(\boldsymbol y\mid\ell)\right)\nonumber\\
&=K\boldsymbol\Delta_{\ell,\ell'}^{\mathsf T}\boldsymbol{\Sigma}^{-1} \boldsymbol\Delta_{\ell,\ell'}=K\sum_{m=1}^M\frac{\Delta_{\ell,\ell',m}^2}{{\sigma}_{m}^2}
\end{align}  
where $\boldsymbol\Delta_{\ell,\ell'}=[\Delta_{\ell,\ell',1},\ldots,\Delta_{\ell,\ell',M}]^{\mathsf T}$ with $\Delta_{\ell,\ell',m}\triangleq\mu_{\ell,m}-\mu_{\ell',m} $ denotes the inter-class mean difference over the $m$-th feature dimension.
\end{lemma}

\begin{figure}[!h]
    \centering
    \includegraphics[width=0.5\textwidth]{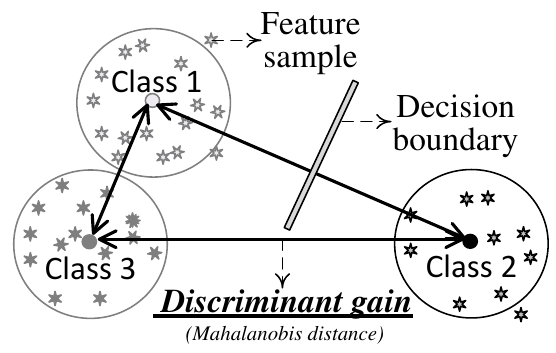}
\caption{Geometric interpretation of the discriminant gain.}
    \label{2Fig2_geometric}
\end{figure}

Based on Lemma~\ref{lemma:GM_aggregation_gain}, we obtain the following
observations:
\begin{enumerate}
    \item A larger DG suppresses the exponential terms in the lower bound of $H\big(\ell \mid \boldsymbol y\big)$. For linear classification, the class average DG $\bar{\mathcal{G}}^{\boldsymbol y}$ quantifies the separability between two classes within the feature space. From a geometric perspective, the DG corresponds to the \emph{Mahalanobis distance} between two classes; see Fig.~\ref{2Fig2_geometric}. 
    \item In binary classification, DG reduces to the \emph{J-divergence}, which is a distance measure commonly used to quantify the detection probability \cite{feres2023over}. Define the average inter-class mean difference as $\bar{\Delta}_{m}\triangleq\left(\frac{2}{L(L-1)}\sum_{\ell=1}^{L}\sum_{\ell'=\ell+1}^{L}\Delta_{\ell,\ell',m}^{2}\right)^{1/2}$. Then,  \eqref{eq:average_DG} can be written as 
    \begin{align}\label{eq:average_DG_nor}
        \bar{\mathcal{G}}^{\boldsymbol y} =K \sum_{m=1}^{M}
\frac{\bar{\Delta}_{m}^{2}}
{{\sigma}_{m}^2}.
    \end{align}
    where ${\bar{\Delta}_{m}^{2}}/{{\sigma}_{m}^2}$ denotes the average DG of dimension-$m$. It is clear that class average DG reflects a well-known a tradeoff between the inter-class separation $\bar{\Delta}_{m}^2$ and the intra-class variance $\sigma_{m}^2$ in Fisher’s linear discriminant analysis \cite{bishop2006pattern}. 
\end{enumerate}

In this work, we address the following question: whether task-aware design always outperforms the classic AirComp design? To facilitate tractable analysis~\cite{chen2024view,wen2023task,wang2026revisiting}, we adopt the class average DG as a task-aware proxy. We first revisit the classical AirComp aggregation, then introduce the task-aware aggregation, and finally compare their resulting structures and performance, with a particular focus on TDM transmission. 
\subsection{Noisy Aggregation Analysis under TDM}\label{sec:TDM}

\subsubsection{Classical Aggregation} The error introduced by AirComp is commonly quantified by the distortion between the ideal received signal in \eqref{Eq:ideal_received_signal} and the aggregated feature vector in \eqref{Eq:aggregation_received_signal}, termed AirComp error, which is given by \cite{cao2020optimized} 
\begin{align}
\label{Eq:AirComp_statistical_error}
\mathcal{D} \!=\!\frac{1}{K^2M}\!\sum_{m=1}^{M}\!\!\Big(\underbrace{ \!\sum_{k=1}^{K}\big|a_m h_{k,m} b_{k,m}\!\!-\!\!1\big|^2\!\nu_{k,m}^2 \!\!+\!|a_m|^2\!\sigma_w^2}_{\mathcal{D}_m}\Big),
\end{align}

where $\mathcal{D}_m$ denotes the distortion of the dimension-$m$ and $\nu_{k,m}^2 \triangleq \mathbb{E}[|{x}_{k,m}|^2]$ denotes the average symbol power \footnote{To keep the paper focused, we assume that the observations $\{\boldsymbol{x}_k\}$ across different devices are approximately uncorrelated. Otherwise, \eqref{Eq:AirComp_statistical_error} serves as an approximate on the AirComp error \cite[Eq.~(10)]{cao2020optimized}. Meanwhile, the average symbol power statistics $\nu_{k,m}^2$ are computed after globally centering all training samples across all classes.}. Under independent per-slot power constraints, AirComp optimization is separable across feature dimensions. Then, the general AirComp optimization problem can be formulated as
\begin{align*}
(\mathrm{P}1) \min_{\{b_{k,m},a_m\}_{m=1}^{M}} 
 \mathcal{D}_m \quad\quad
\text{s.t.}\quad 
|b_{k,m}|^2 \nu_{k,m}^2 \le  P_k, \quad\forall k.
\end{align*}
where $P_{k}$ denotes the power budget of each slot.  The transmit SNR is then defined as $\mathrm{SNR}_{\mathsf{c}}=10\log_{10}\left(\frac{P_{k}}{\sigma_w^2}\right) $ \cite{chen2024view}. Existing work has shown that the optimal solution to $(\mathrm{P}1)$ admits a threshold-based structure \cite{cao2020optimized}; therefore, the detailed derivation is omitted here.

\subsubsection{Task-Aware Aggregation}Classical aggregation aims to minimize the AirComp error, which is not fully aligned with our ultimate objective of improving task performance. Motivated by the observation in Lemma~\ref{lemma:GM_aggregation_gain}, we next aim to maximize the DG of the received signal, termed task-aware aggregation. The DG of the aggregated feature $\boldsymbol y^{\sf t}$ in \eqref{eq:TDM_channel}, is expressed as
 \begin{align}\label{eq:class_average_DG_TDM}
\bar{\mathcal{G}}^{\boldsymbol y^{\sf t}}=\sum_{m=1}^{M}\underbrace{\frac{\big|\sum_{k=1}^{K} h_{k,m}b_{k,m}\big|^2\bar{\Delta}_{m}^2}{\sum_{k=1}^{K} \!\!|h_{k,m} b_{k,m}|^2{\sigma}_{m}^2\!+\!\sigma_w^2}}_{\bar{\mathcal{G}}^{ y_m^{\sf t}}}.
\end{align}
For an independent time slot $m$, the corresponding optimization problem here is formulated as
\begin{align*}
\mathrm{(P2)}
\max_{\{b_{k,m},a_m\}_{m=1}^{M}}
\bar{\mathcal{G}}^{ y_m^{\sf t}} \quad\quad
\text{s.t.} \quad |b_{k,m}|^2\nu_{k,m}^2 \le  P_k, \quad \forall k.
\end{align*} 
It can be observed that the DG is independent of the aggregation coefficient $a_m$. Hence, the optimization of $\mathrm{(P2)}$ can be reduced to the design of the transmit coefficients ${b_{k,m}}$. Since the channel is slow fading, the channel remains unchanged over the $M$ time slots, i.e., ${h}_{k,m} = {h}_k, \, \forall m$. We assume phase alignment at the transmitter by setting $b_{k,m} = \tilde{b}_{k,m}e^{-j\angle h_{k}}$ with acquired CSI, where $\tilde{b}_{k,m}$ denotes the transmit amplitude and $\angle h_{k}$ symbolizes the phase of the channel $h_{k}$. The channel amplitude is denoted by $\tilde{h}_{k} \triangleq |h_{k}|$. Then, the optimal solution to $(\mathrm{P}2)$ is summarized as follows.

\begin{proposition}[Task-Aware Aggregation under TDM]
\label{proposition:dg_optimal_TDM}
The optimal solution to $(\mathrm{P}2)$ exhibits a threshold-based structure, given by
\begin{equation}\label{eq:p3_solution} 
\tilde b_{k,m}^\star
=\min\left\{
\frac{\sqrt{P_k}}{\nu_{k,m}},
\frac{\tau^\star}{\tilde h_k{\sigma}_{m}^2}
\right\},
\quad \forall k.
\end{equation}
where $\tau^\star$ denotes the optimal threshold.
\end{proposition}
\begin{proof}
See Appendix \ref{appendix:Proof of proposition1}.
\end{proof}
\begin{figure}[!h]
	\centering
	\begin{minipage}{0.3\textwidth}
		{\includegraphics[width=\textwidth]
			{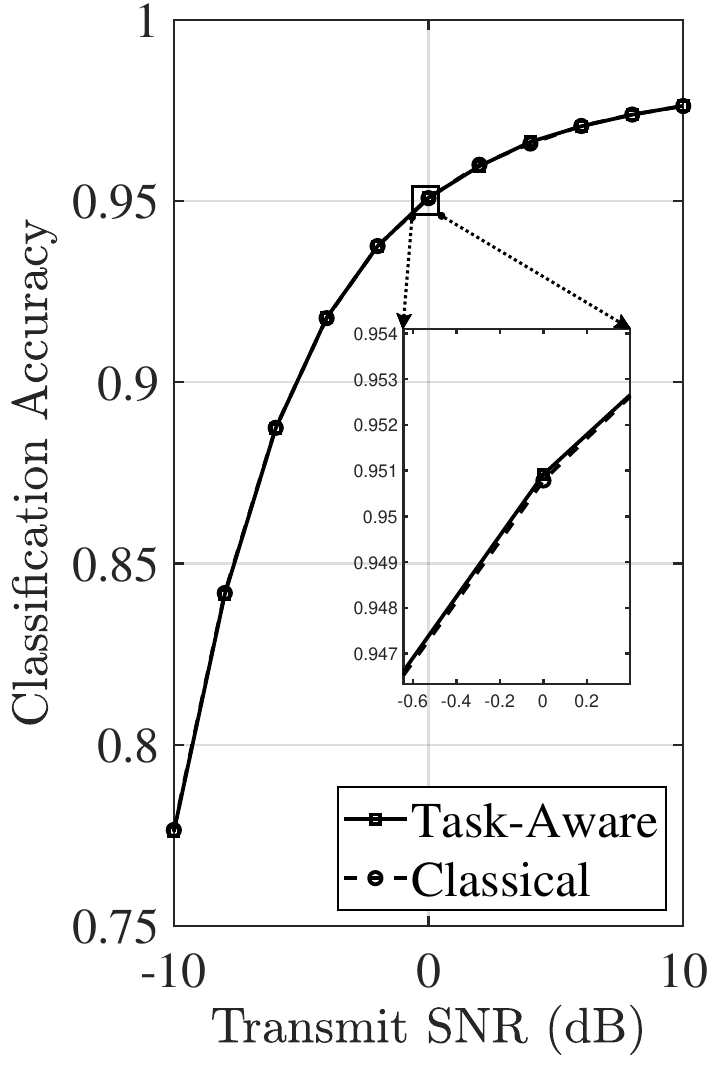}}
		\caption*{(a)}\label{Fig_P_MSE_vs_DG_GMM_TDM_ACC}        
	\end{minipage}
		\begin{minipage}{0.3\textwidth}
		{\includegraphics[width=\textwidth]
			{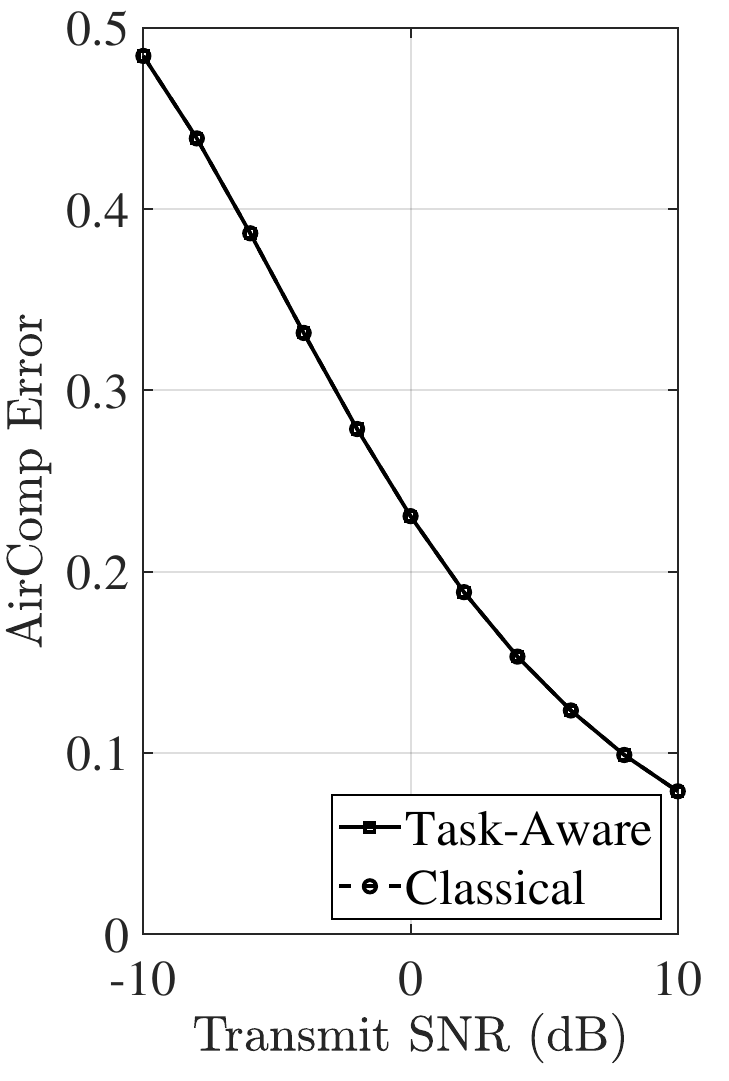}}
		\caption*{(b)}\label{Fig_P_MSE_vs_DG_GMM_TDM_MSR}
	\end{minipage}
\caption{Performance comparison between the task-aware and classical aggregation versus transmit SNRs under TDM transmission: (a) classification accuracy and (b) AirComp error.}
\label{Fig_P_acc_snr_TDM}
\end{figure}

\subsubsection{Numerical Comparison}
Proposition~\ref{proposition:dg_optimal_TDM} reveals that task-aware aggregation exhibits a similar threshold-based structure with classic aggregation \cite[Th.~1]{cao2020optimized}. This result is validated by the numerical example in Fig.~\ref{Fig_P_acc_snr_TDM}, where a GM classification task is considered with $K=10$, $L=20$, $M=30$ and linear classifier in \eqref{eq:map_classifier}. The main observations are summarized as follows:

\begin{enumerate}
\item As illustrated in Fig.~\ref{Fig_P_acc_snr_TDM}, task-aware aggregation and classical aggregation achieve identical performance in terms of classification accuracy and AirComp error. The underlying reason is that, different feature dimensions are independently optimized under identical channel realization. Although the discriminability prior $\{{\bar{\Delta}_{m}^{2}}/{{\sigma}_{m}^2}\}$ is known, the independent per-slot power constraints prevent resource reallocation across feature dimensions. 

 \item  The identical channel realization across time slots eliminates the opportunity to jointly match feature importance with channel quality. In this regard, irrespective of the adopted optimization objective, the resulting solution exhibits the same structural form.
\end{enumerate}

We notice that the discriminability prior $\{{\bar{\Delta}_{m}^{2}}/{{\sigma}_{m}^2}\}$ is heterogeneous across feature dimensions. To better utilize this prior, it is beneficial for the channel \emph{degrees of freedom} (DoFs) to support corresponding multiplexing, enabling non-uniform resource allocation that aligns with feature importance. Motivated by this observation, we next extend this analysis to the FDM transmission.

\section{Task-Aware FDM Transmission for AirComp}\label{sec:Task-Aware}
In this section, we analyze the aggregation performance under FDM transmission. Our analysis consists of three steps. First, we formulate and solve the classical AirComp and task-aware AirComp problems. Second, we compare the two problems in terms of how they balance inter-class separation and intra-class variance. Finally, we extend the analysis to the practical case where the receiver is equipped with multiple antennas.
\subsection{FDM Transmission}\label{sec:aggregation_analysis_fdm}
\subsubsection{Classical Aggregation}
Different from TDM transmission, each feature element here is transmitted over different subcarriers rather than different time slots; see Fig.~\ref{5_Fig5_TDM_FDM}(b).
Due to the frequency-selective channel, a joint power constraint across all subcarriers should be considered. Similar to the TDM transmission, phase alignment is performed as
$b_{k,m} = \tilde{b}_{k,m} e^{-j\angle h_{k,m}}$. We further define $\tilde{h}_{k,m} \triangleq |h_{k,m}|$ and $\tilde{a}_{m} \triangleq |a_m|$. Under independent per-slot power constraints, the corresponding optimization problem can be formulated as
\begin{equation*}
(\mathrm{P}3) \!\!\!\!\min_{\{\tilde{b}_{k,m},\tilde{a}_m\}_{m=1}^{M}} \sum_{m=1}^{M} \mathcal{D}_m \quad
\text{s.t.}  \sum_{m=1}^M\tilde{b}_{k,m}^2\nu_{k,m}^2 \le P_k, \; \forall k.
\end{equation*}
Previous studies have shown that the optimal transmit coefficient $\tilde{b}_{k,m}$ for $(\mathrm{P}3)$ follows a regularized channel-inversion structure, whereas the optimal receive filter $\tilde{a}_{m}$ takes the form of Wiener filter~\cite{chen2023over,xie2023optimal}.

\subsubsection{Task-Aware Aggregation}
We next aim to maximize the DG of the received signal under FDM transmission, which leads to a task-aware FDM mechanism. As shown later, this mechanism allocates more resources to feature dimensions with higher discriminative importance. The corresponding optimization problem can be formulated as
\begin{align*}
\mathrm{(P4)}
\max_{\{\tilde{b}_{k,m},\tilde{a}_m\}_{m=1}^{M}}\quad
\bar{\mathcal{G}}^{\boldsymbol y^{\sf f}} \quad
\text{s.t.}
\sum_{m=1}^M\tilde{b}_{k,m}^2\nu_{k,m}^2 \le P_k, \quad \forall k.\nonumber
\end{align*} 
where 
\begin{align}\label{eq:dg_fdm}
    \bar{\mathcal{G}}^{\boldsymbol y^{\sf f}}=\sum_{m=1}^{M}\underbrace{ \frac{\big(\sum_{k=1}^{K} \tilde{h}_{k,m}\tilde{b}_{k,m}\big)^2\bar{\Delta}_{m}^2}{\sum_{k=1}^{K} \!\!\tilde{h}_{k,m}^2 \tilde{b}_{k,m}^2{\sigma}_{m}^2\!+\!\sigma_w^2}}_{\bar{\mathcal{G}}^{ y_m^{\sf f}}}.
\end{align}

$(\mathrm{P}4)$ is non-convex due to the coupling between ${\tilde{b}_{k,m}}$ across devices and subcarriers. It can be verified that $(\mathrm{P}4)$ satisfies the time-sharing condition \cite{yu2006dual}. Next, $(\mathrm{P}4)$ can be optimally solved by the following proposition.

\begin{proposition}[Task-Aware Aggregation under FDM]\label{proposition:decision_optimal_fdm}
    $(\mathrm{P}4)$ can be solved using the Lagrange duality method. The resulting dual-decomposition algorithm, together with its convergence and complexity analysis, is presented in Appendix~\ref{appendix:p2_lagrange_duality} and summarized in Algorithm~\ref{alg:dual_P2_dg}.
\end{proposition}

\begin{algorithm}[!t]
\caption{Dual-Decomposition for $(\mathrm{P}4)$}
\label{alg:dual_P2_dg}
\begin{algorithmic}[1]
    \STATE \textbf{Input:} $\{{\sigma}_{m}^2\}$, $\{\bar{\Delta}_m\}$, $\sigma_w^2$, $\{\nu_{k,m}^2\}$, $\{P_k\}$, $\varepsilon_\lambda$, and $\varepsilon_p$.
    \STATE \textbf{Initialize:} generate $\lambda_k^{(0)}>0$, $z_m^{(0)}>0$; set iteration index $t\gets 0$.
    \REPEAT
        \STATE \textbf{(Update per-dimension auxiliary)} For each $m$, solve \eqref{eq:consistency_condition_dg} with given $\{\lambda_k^{(t)}\}$ to obtain $z_m^{(t+1)}$;
        \STATE \textbf{(Closed-form primal update)} For each $k$ and $m$, compute $\tilde{b}_{k,m}^{(t+1)}$ using \eqref{eq:optimal_b_k_m_dg};
        \STATE \textbf{(Update per-device dual variable)} For each $k$, update $\lambda_k$ using the subgradient $P_k-\sum_{m=1}^{M}\nu_{k,m}^2|\tilde{b}_{k,m}^{(t+1)}|^2$;
        \STATE $t \gets t+1$;
    \UNTIL{$\max_k |\lambda_k^{(t)}-\lambda_k^{(t-1)}|\le\varepsilon_\lambda$ \textbf{and} power violations are below $\varepsilon_p$}
    \STATE \textbf{Output:} $\tilde{b}_{k,m}^{\star}=\tilde{b}_{k,m}^{(t)}$.
\end{algorithmic}
\end{algorithm}

\subsubsection{Numerical Comparison}
\begin{figure}[!h]
	\centering
	\begin{minipage}{0.3\textwidth}
		{\includegraphics[width=\textwidth]
			{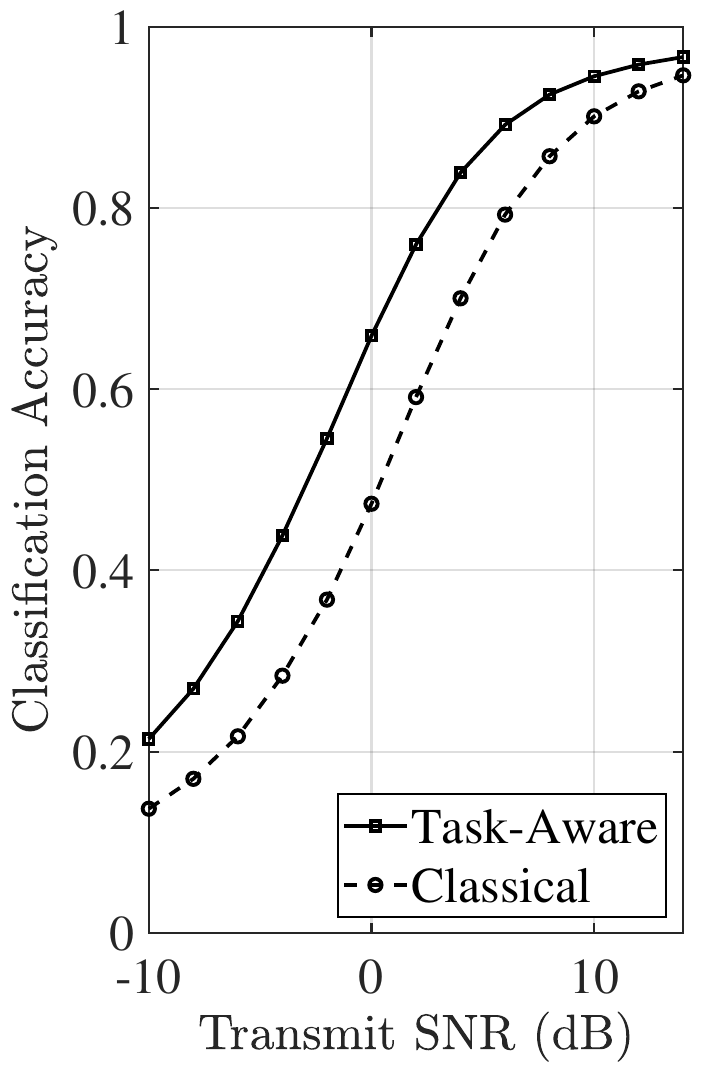}}
		\caption*{(a)}\label{Fig_P_MSE_vs_DG_GMM_TDM_ACC}        
	\end{minipage}
		\begin{minipage}{0.3\textwidth}
		{\includegraphics[width=\textwidth]
			{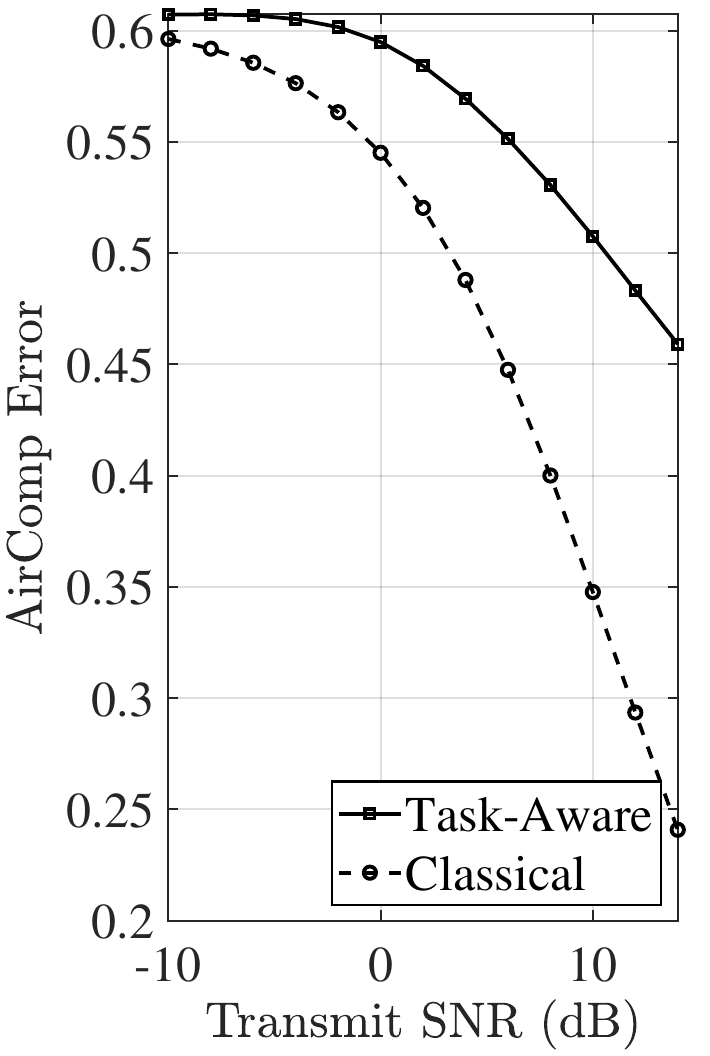}}
		\caption*{(b)}\label{Fig_P_MSE_vs_DG_GMM_TDM_MSR}
	\end{minipage}
    \caption{Performance comparison between the task-aware aggregation and classical aggregation versus transmit SNRs under FDM transmission: (a) classification accuracy and (b) AirComp error.}
    \label{Fig_P_acc_snr_FDM}
\end{figure}

Proposition~\ref{proposition:decision_optimal_fdm} is validated by the numerical example in Fig.~\ref{Fig_P_acc_snr_FDM}; the simulation settings are the same as the TDM transmission. It is observed that task-aware aggregation achieves better classification accuracy than the Classical aggregation [Fig.~\ref{Fig_P_acc_snr_FDM}(a)], despite yielding a higher AirComp error [Fig.~\ref{Fig_P_acc_snr_FDM}(b)]. To gain deeper insight into this phenomenon, we provide the following analysis.

\subsection{Analysis on Task-Aware FDM: Tradeoff}\label{sec:Discussion_Aware}

We compare the classical aggregation and the task-aware aggregation in terms of how they balance inter-class separation and intra-class variance, see the following corollary. 

\begin{corollary}\label{corollary:mse_equivalent_objective}
Minimizing the AirComp error in \eqref{Eq:AirComp_statistical_error} is equivalent to maximizing the following error-related utility:
\begin{equation}\label{eq:equivalentMSE}
\mathcal U_m
=
\frac{\left(\sum_{k=1}^{K}\tilde h_{k,m}\tilde b_{k,m}\nu_{k,m}^2\right)^2}
{\sum_{k=1}^{K}\tilde h_{k,m}^2\tilde b_{k,m}^2\nu_{k,m}^2+\sigma_w^2}.
\end{equation}
\end{corollary}
\begin{proof}
    See Appendix~\ref{appendix:mse_equivalent_obj}.
\end{proof}

Comparing \eqref{eq:equivalentMSE} with \eqref{eq:dg_fdm}, the main observations are given as follows:

\begin{enumerate}
    \item The classical aggregation relies on the average symbol power $\nu_{k,m}^2$ in \eqref{eq:equivalentMSE}, while the task-aware aggregation explicitly exploits discriminability prior ${\bar{\Delta}_{m}^{2}}/{{\sigma}_{m}^2}$ in \eqref{eq:dg_fdm}.
    \item The task-aware aggregation performs non-uniform power allocation across subcarriers by jointly accounting for ${\bar{\Delta}_{m}^{2}}/{{\sigma}_{m}^2}$,  thereby striking a tradeoff between the inter-class separation $\bar{\Delta}_{m}^2$ and the intra-class variance $\sigma_{m}^2$.
\end{enumerate}

We now revisit the TDM case discussed in Section~\ref{sec:TDM}. We previously claimed that different feature dimensions are independently optimized under quasi-static time slots, thereby preventing task-aware resource allocation. Actually, TDM can also realize task-aware resource allocation under a cross-slot energy constraint. Under slow fading, such TDM design can allocate energy according to feature importance but cannot exploit temporal channel heterogeneity. Under fast fading with only causal CSI, it becomes a sequential decision problem, while acquiring timely CSI is challenging. One potential approach is to predict the timely CSI, e.g., using LSTM-based methods \cite{peng2020lstm}. This comes at the cost of high computational complexity and implementation difficulty. CSI acquisition and its impact on the proposed task-aware FDM design will be discussed in the next section.

\subsection{Analysis on Task-Aware FDM: Multiple-Antenna Case}\label{sec:MIMO}

We end this subsection by extending the preceding analysis to the multiple-antenna case, where the AP is equipped with $N_{\sf r}$ receive antennas. Accordingly, \eqref{eq:dg_fdm} can be rewritten as
\begin{equation}\label{eq:dg_fdm_simo} {\bar{\mathcal{G}}}^{\boldsymbol y^{\sf f}} = \sum_{m=1}^{M} \frac{ \left| \sum_{k=1}^{K} \boldsymbol{a}_m^{\mathsf H} \boldsymbol{h}_{k,m} \tilde{b}_{k,m} \right|^2 \bar{\Delta}_{m}^{2} }{ \sum_{k=1}^{K} \left| \boldsymbol{a}_m^{\mathsf H} \boldsymbol{h}_{k,m} \tilde{b}_{k,m} \right|^2 {\sigma}_{m}^2 + \sigma_w^2\|\boldsymbol{a}_m\|^2 },
\end{equation}
where $\boldsymbol{h}_{k,m}\in\mathbb{C}^{N_{\sf r}}$ denote the channel vector from device $k$ to the AP over the $m$-th subcarrier and $\boldsymbol{a}_m\in\mathbb{C}^{N_{\sf r}}$ denotes the corresponding receive combining vector. Under independent Rayleigh fading, we have $\boldsymbol{h}_{k,m}=\sqrt{\varrho_{k}}\boldsymbol{g}_{k,m}$, where $\varrho_k$ and $\boldsymbol{g}_{k,m}\sim\mathcal{CN}(\boldsymbol{0},\mathbf I_{N_{\sf r}})$ denote the large-scale and small-scale fading, respectively.

Then, we revisit $\mathrm{(P4)}$, which is non-convex due to the coupling between the transmit coefficients $\tilde{b}_{k,m}$ and the receive combining vectors $\boldsymbol{a}_m$. A standard solution is to alternately optimize the transmitter and receiver. However, this approach offers limited analytical insight. We thus establish the following lemma to characterize the impact of multi-antenna diversity on the DG performance under FDM transmission.

\begin{lemma}[Receive-Antenna Scaling Law]
\label{lemma:receive_antenna_scaling}For a fixed $K$, as $N_{\sf r}\rightarrow\infty$, each device can scale its transmit power as $\tilde{b}_{k,m}^2= \bar b_{k,m}^2/N_{\sf r}$ while almost surely maintaining the same asymptotic DG as
 \begin{equation}\label{eq:dg_simo_asymptotic}
\lim_{N_{\sf r}\to\infty}\bar{\mathcal{G}}^{\boldsymbol y^{\sf f}}
=\sum_{m=1}^{M}\frac{\bar{\Delta}_{m}^{2}
\left|\sum_{k=1}^{K}\varrho_{k}\bar b_{k,m}
\right|^{2}}{\sum_{k=1}^{K}\varrho_{k}^{2}|\bar b_{k,m}|^{2}{\sigma}_{m}^2+\sigma_w^2\sum_{k=1}^{K}\varrho_{k}}.
    \end{equation}
where $\bar b_{k,m}^2$ denotes the reference power.
\end{lemma}

\begin{proof}
See Appendix~\ref{appendix:receive_antenna_scaling}.
\end{proof}
\begin{figure}[!h]
	\centering
	\begin{minipage}{0.3\textwidth}
		{\includegraphics[width=\textwidth]
			{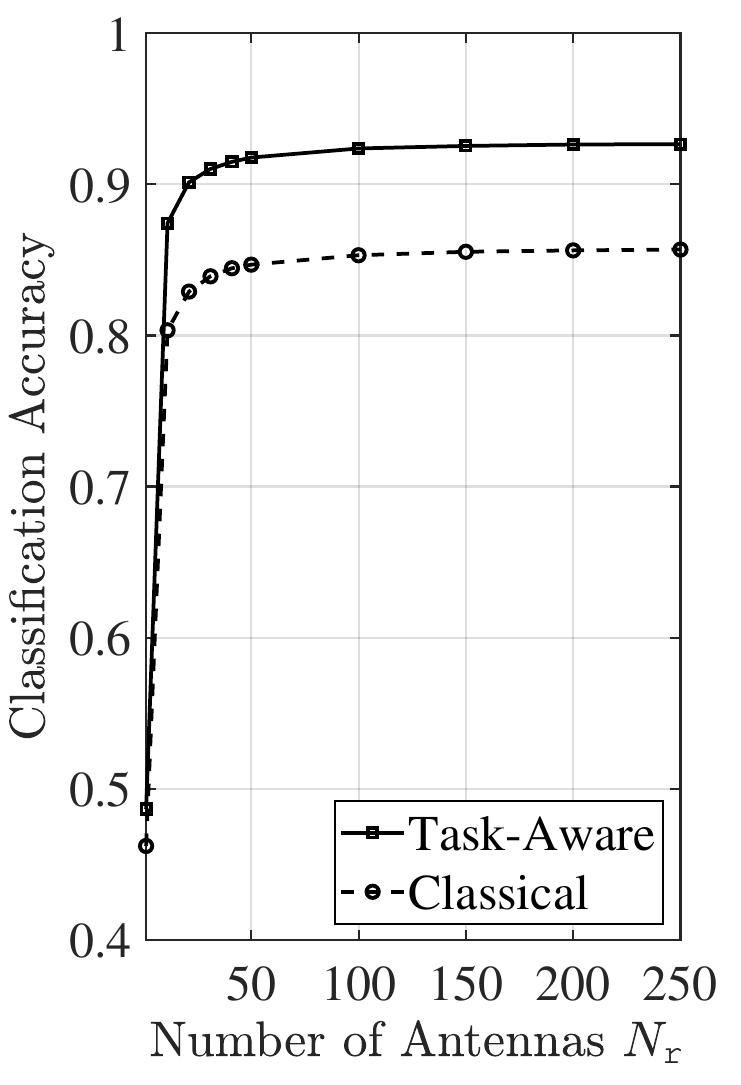}}
		\caption*{(a)}\label{Fig_P_MSE_vs_DG_GMM_TDM_ACC}        
	\end{minipage}
		\begin{minipage}{0.3\textwidth}
		{\includegraphics[width=\textwidth]
			{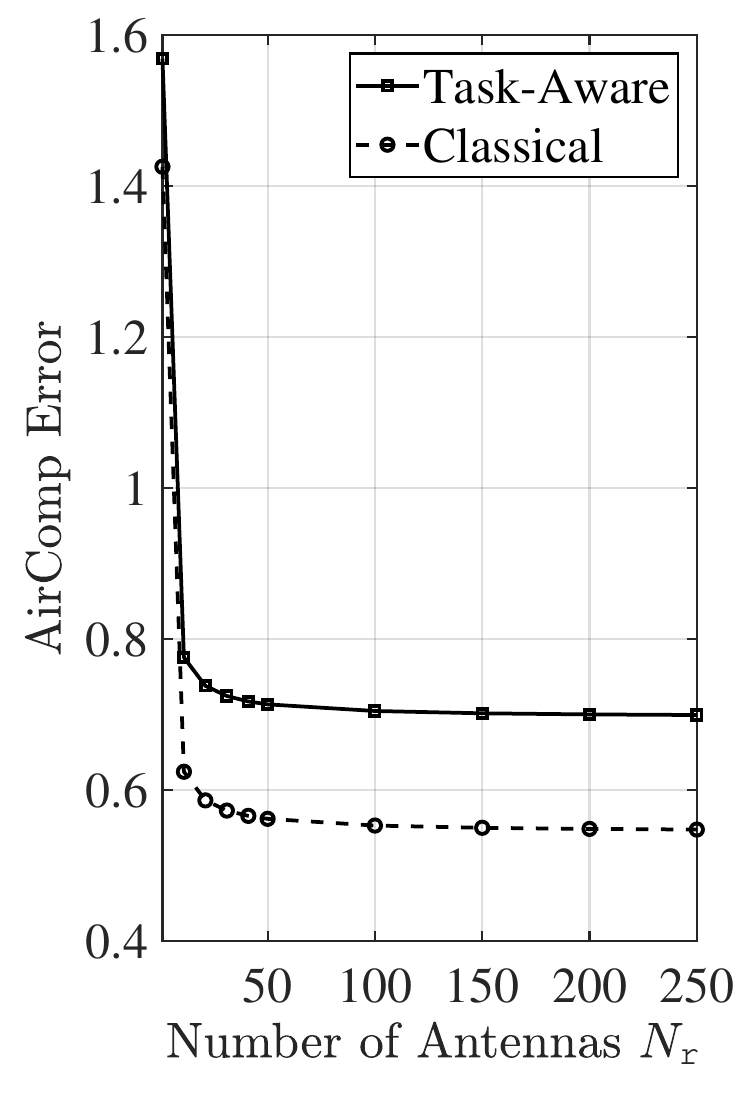}}
		\caption*{(b)}\label{Fig_P_MSE_vs_DG_GMM_TDM_MSR}
	\end{minipage}
	\caption{
    Performance versus the number of receive antennas: (a) classification accuracy and (b) AirComp error.}\label{Fig_P5_acc_antenna}
\end{figure}

Lemma~\ref{lemma:receive_antenna_scaling} shows that:
\begin{enumerate}
    \item Deploying more antennas at the AP allows the devices to reduce their transmit power without degrading the DG performance, which shares a similar spirit with massive MIMO in wireless cellular communications \cite{ngo2013energy}. This leads to prolonged battery life of each device.
    \item As $N_{\sf r}\to\infty$, channel hardening and asymptotic orthogonality imply that the asymptotic DG depends only on the statistical CSI $\varrho_k$. The transceiver design in $(\mathrm{P}4)$ can still be solved by Algorithm~\ref{alg:dual_P2_dg} with statistical CSI.
\end{enumerate}
 Lemma~\ref{lemma:receive_antenna_scaling} is verified in Fig.~\ref{Fig_P5_acc_antenna}, where the performance approaches the asymptotic bound as the number of antennas $N_{\sf r}$ increases. With Lemma~\ref{lemma:receive_antenna_scaling} in hand, we have the following corollary. 

\begin{corollary}[Upper Bound of Asymptotic DG]
\label{corollary:sensing_noise_limited_bound}
The asymptotic DG in \eqref{eq:dg_simo_asymptotic} is upper bounded by
\begin{equation}\label{eq:dg_simo_upper_bound}
\lim_{N_{\sf r}\to\infty}{\bar{\mathcal{G}}}^{\boldsymbol y^{\sf f}}
\le
K\sum_{m=1}^{M}\frac{\bar{\Delta}_{m}^{2}}{{\sigma}_{m}^2}.
\end{equation}
\end{corollary}

\begin{proof}
For the $m$-th feature dimension, we have
\begin{align}
&{\bar{\mathcal{G}}}^{y_m^{\sf f}}
=\frac{\bar{\Delta}_{m}^{2}\left|\sum_{k=1}^{K}\varrho_k\bar b_{k,m}\right|^{2}}{\sum_{k=1}^{K}\varrho_k^{2}|\bar b_{k,m}|^{2}\sigma_m^{2}
+\sigma_w^{2}\sum_{k=1}^{K}\varrho_k}\nonumber\\
&\overset{(a)}{\leq}
\frac{\bar{\Delta}_{m}^{2}}{\sigma_m^{2}}
\frac{K\sum_{k=1}^{K}\varrho_k^{2}|\bar b_{k,m}|^{2}
}{\sum_{k=1}^{K}\varrho_k^{2}|\bar b_{k,m}|^{2}
+\frac{\sigma_w^{2}}{\sigma_m^{2}}
\sum_{k=1}^{K}\varrho_k}\overset{(b)}{\leq}
\frac{K\bar{\Delta}_{m}^{2}}{\sigma_m^{2}},
\label{eq:single_feature_dg_upper_bound}
\end{align}
where (a) follows from the Cauchy--Schwarz inequality
and (b) follows from $\frac{\sum_{k=1}^{K}\varrho_k^{2}|\bar b_{k,m}|^{2}
}{\sum_{k=1}^{K}\varrho_k^{2}|\bar b_{k,m}|^{2}+
\frac{\sigma_w^{2}}{\sigma_m^{2}}
\sum_{k=1}^{K}\varrho_k}\leq 1$. Summing \eqref{eq:single_feature_dg_upper_bound}
over all feature dimensions yields \eqref{eq:dg_simo_upper_bound}.
\end{proof}

Corollary~\ref{corollary:sensing_noise_limited_bound} shows that, as $N_{\sf r}\to\infty$, the upper bound of the asymptotic DG coincides with noise-free aggregation in \eqref{eq:average_DG_nor}, which is limited solely by sensing noise rather than communication noise.

\section{CSI Acquisition and Nonlinear DNN Classification} \label{sec:csi}

The preceding analysis was developed under the assumption of perfect CSI and linear classification.
In this section, we first investigate the CSI acquisition procedure and examine the impact of imperfect CSI. Building on the insights derived from the linear classification analysis, we then develop a task-aware FDM mechanism for practical nonlinear classification tasks based on DNNs.

\begin{figure}[!h]
    \centering
    \includegraphics[width=0.6\textwidth]{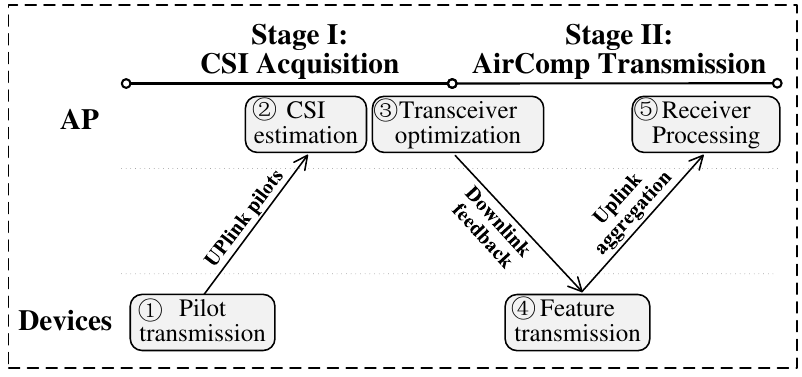}
\caption{AirComp aggregation workflow.}
    \label{4_Fig4_CSI}
\end{figure}
\subsection{CSI Acquisition}
To realize the task-aware FDM, the uplink CSI $\{\boldsymbol{h}_k\}_{k=1}^K$ has to be estimated at the AP. The standard way of doing this is to use uplink pilots. As illustrated in Fig.~\ref{4_Fig4_CSI}, a complete AirComp aggregation workflow consists of two stages: \textbf{Stage~I} CSI acquisition and \textbf{Stage~II} AirComp transmission. Stage~II has been detailed in the previous sections; we thus focus on Stage~I here. For each device, the uplink CSI $\boldsymbol{h}_k$ contains $M$ subcarriers. Estimating $M$ subcarriers may incur much pilot overhead; we thus exploit the fact that subcarriers within the coherence bandwidth exhibit similar fading \cite{negi1998pilot}. Hence, pilots can be transmitted only over $N_{\sf p}$ representative subcarriers, where $N_{\sf p} \ll M$ is determined by the channel delay spread \cite{tse2005fundamentals}. Upon receiving the pilots, the AP can estimate the CSI using classic estimation methods, such as LMMSE \cite{kay1993fundamentals}. The full CSI vector across $M$ subcarriers is reconstructed via frequency-domain interpolation. An illustrative example is given as follows. 

\begin{example}[CSI Acquisition for Task-Aware FDM] The pilot vector used by device $k$ is denoted by $\boldsymbol{\omega}_{k} =[\omega_{k,1},\omega_{k,2},\ldots,\omega_{k,N_{\sf p}}]^{\mathsf T}\in\mathbb{C}^{N_{\sf p}}$, with $|\omega_{k,n}|^2=1$. The corresponding received training signal is 
\begin{equation*}
\boldsymbol{y}_{k}^{\sf p} = \sqrt{p_{k}^{\sf p}}\, \boldsymbol{\omega}_{k}\odot\boldsymbol{h}_{k}^{\sf p} + \boldsymbol{w}_{k}^{\sf p},
\end{equation*} 
where $p_{k}^{\sf p}$ denotes the pilot transmit power, $\boldsymbol{h}_{k}^{\sf p} = [h_{k,1},h_{k,2},\ldots,h_{k,N_{\sf p}}]^{\mathsf T}\in\mathbb{C}^{N_{\sf p}}$ and $\boldsymbol{w}_{k}^{\sf p}\sim \mathcal{CN}(\boldsymbol{0},\sigma_{w}^{2}\mathbf{I}_{N_{\sf p}})$. Assuming $\boldsymbol{h}_{k}^{\sf p}\sim \mathcal{CN}(\boldsymbol{0},\varrho_k\mathbf I_{N_{\sf p}})$, the LMMSE estimate of $\boldsymbol{h}_{k}^{\sf p}$ is given by \cite{kay1993fundamentals} 
\begin{equation}\label{eq:received_csi}
\hat{\boldsymbol{h}}_{k}^{\sf p}
=\frac{
\sqrt{p_{k}^{\sf p}}\varrho_k
}{p_{k}^{\sf p}\varrho_k+\sigma_{w}^{2}
}{\rm diag}(\boldsymbol{\omega}_{k})^{\mathsf H}
\boldsymbol{y}_{k}^{\sf p}.
\end{equation} 
Then, the full CSI vector over all $M$ subcarriers is reconstructed as
\begin{equation}\label{eq:full_interpolation}
\hat{\boldsymbol h}_k
=\boldsymbol{\varpi}_k
\hat{\boldsymbol h}_k^{\sf p},
\end{equation}
where $\boldsymbol{\varpi}_k\in\mathbb{C}^{M\times N_{\sf p}}$ denotes the
frequency-domain interpolation matrix. 
The estimated CSI $\hat{\boldsymbol{h}}_{k}$ is then used for the subsequent transceiver design in Stage~II. \end{example}

After channel estimation at the AP, the transceiver can be
optimized according to Algorithm~\ref{alg:dual_P2_dg}, yielding
$\{\tilde{b}_{k,m}^{\star}\}$. A straightforward implementation is to feedback all $\{\tilde{b}_{k,m}^{\star}\}$ to each device [see Fig.~\ref{4_Fig4_CSI}]; however, this incurs a downlink overhead proportional to $KM$. To reduce this overhead, the AP can instead broadcast the auxiliary
variables $\{z_m^\star\}_{m=1}^{M}$ and
separately feedback the dual variable $\lambda_k^\star$ in \eqref{eq:optimal_b_k_m_dg} to device $k$. Then, each device locally computes $\tilde{b}_{k,m}^{\star}$ according to \eqref{eq:optimal_b_k_m_dg}. The feedback overhead is thus reduced to
$\mathcal{O}(M+K)$. This implementation requires each device to acquire its own CSI. Specifically, each device can operate in the \emph{time-division duplex} (TDD) mode and exploit channel reciprocity to obtain the CSI \cite[Ch.~5]{tse2005fundamentals}. The time interval between uplink and downlink transmissions satisfies $\Delta t_{\mathrm{TDD}} < T_{\mathrm{cd}}$, ensuring that the channel remains constant during the TDD cycle. 

\subsection{Analysis on Imperfect CSI: Regularization}
Any CSI acquisition method inevitably introduces CSI estimation error. Based on the LMMSE estimate in \eqref{eq:received_csi}, we let $\boldsymbol{g}_{k,m}^{\mathsf{H}}$ be the $m$-th row of the interpolation matrix $\boldsymbol{\varpi}_k$ and define $\rho_{k,m} \triangleq \|\boldsymbol{g}_{k,m}\|^2$. The imperfect CSI model is summarized in the following proposition.

\begin{proposition}[Imperfect CSI Error Model]
\label{proposition:per_subcarrier_csi_error_model}
The estimated CSI can be modeled as the summation of the true CSI $h_{k,m}$ and estimation error $e_{k,m}$
\begin{equation}\label{eq:csi_error_model}
\hat h_{k,m}=h_{k,m}+e_{k,m},
\end{equation}
where $\hat h_{k,m}$ and $e_{k,m}$ are distributed as
\begin{align}
\hat h_{k,m}&\sim\mathcal{CN}\left(0,\varpi_k\rho_{k,m}\right),\quad
\varpi_k=\frac{p_k^{\sf p}\varrho_k^2}{p_k^{\sf p}\varrho_k+\sigma_w^2},\label{eq:es_k}\\
e_{k,m}&\sim\mathcal{CN}\left(0,\kappa_k\rho_{k,m}\right),\quad
\kappa_k=\frac{\varrho_k\sigma_w^2}{p_k^{\sf p}\varrho_k+\sigma_w^2}.\label{eq:es_e}
\end{align}
\end{proposition}

\begin{proof}
See Appendix~\ref{appendix:imperfect_csi_model}.
\end{proof}
Clearly,  the estimation error $e_{k,m}$ in \eqref{eq:es_e} is jointly determined by three factors: (i) the large-scale fading and receiver noise, captured by $\varrho_k$ and $\sigma_w^2$; (ii) the pilot transmit power $p_k^{\sf p}$; and (iii) the interpolation $\rho_{k,m}$.

Based on the imperfect CSI model, $(\mathrm{P}4)$ can be reformulated by first deriving the corresponding DG. Let $b_{k,m} = \tilde{b}_{k,m}e^{-j\angle \hat h_{k,m}}$ with imperfect CSI $\hat h_{k,m}$. Then, the DG is given by
\begin{align}
\label{eq:dg_fdm_imperfect_csi}
{\bar{\mathcal{G}}}^{\boldsymbol y^{\sf f}}
=\sum_{m=1}^{M}
\frac{\left(\sum_{k=1}^{K}|\hat{h}_{k,m}|\tilde b_{k,m}\right)^2
\bar{\Delta}_{m}^{2}}{
\sum_{k=1}^{K}
\left(|\hat{h}_{k,m}|^2+\epsilon_{k,m}^2\right)
\tilde b_{k,m}^{2}{\sigma}_{m}^2
+\sigma_w^2},
\end{align}
where $\epsilon_{k,m}^2 \triangleq \kappa_k\rho_{k,m}$. Comparing \eqref{eq:dg_fdm_imperfect_csi} with \eqref{eq:dg_fdm}, the imperfect CSI introduces an additional regularization term $\epsilon_{k,m}^2$ in the denominator. This problem follows the same solution procedure as Algorithm~\ref{alg:dual_P2_dg}, except that \eqref{eq:optimal_b_k_m_dg} is
modified as
\begin{equation*}
\tilde b_{k,m}^{\star}(\lambda_k,z_m)
=
\frac{\bar{\Delta}_{m}^{2}|\hat h_{k,m}|z_m}
{\lambda_k\nu_{k,m}^{2}
+\bar{\Delta}_{m}^{2}{\sigma}_{m}^2
\left(|\hat h_{k,m}|^2+\epsilon_{k,m}^2\right)z_m^{2}},
\end{equation*}
where the prior-guided structure is retained, and the CSI error $\epsilon_{k,m}^2$ acts as a regularization term. 


\subsection{Analysis on Imperfect CSI: Scaling}
Accordingly, the preceding analysis can be extended to the multiple-antenna case.  The estimated channel in \eqref{eq:csi_error_model} can be rewritten as
\begin{equation*}
\hat{\boldsymbol h}_{k,m}=\boldsymbol h_{k,m}+\boldsymbol e_{k,m},
\end{equation*}
where $\hat{\boldsymbol h}_{k,m}$ and $\boldsymbol e_{k,m}$ are
distributed as
\begin{equation*}
\hat{\boldsymbol h}_{k,m}\sim
\mathcal{CN}\left(\boldsymbol 0,\varpi_k\rho_{k,m}\mathbf I_{N_{\sf r}}\right),
\quad
\boldsymbol e_{k,m}\sim
\mathcal{CN}\left(\boldsymbol 0,\kappa_k\rho_{k,m}\mathbf I_{N_{\sf r}}\right).
\end{equation*}
Then, \eqref{eq:dg_fdm_simo} can be rewritten as
\begin{equation}\label{eq:dg_fdm_simo_ic} 
{\bar{\mathcal{G}}}^{\boldsymbol y^{\sf f}}
\!\!\!=\!\!\!\sum_{m=1}^{M}\!\!\frac{\left|\sum_{k=1}^{K}{\boldsymbol a}_m^{\mathsf H}\hat{\boldsymbol h}_{k,m}\tilde b_{k,m}\right|^2\bar{\Delta}_{m}^{2}}{\sum_{k=1}^{K}\!\!\left(\!\left|{\boldsymbol a}_m^{\mathsf H}\hat{\boldsymbol h}_{k,m}\right|^2\!\!\!+\!\!\epsilon_{k,m}^2\|{\boldsymbol a}_m\|^2
\!\!\right)\tilde b_{k,m}^{2}{\sigma}_{m}^2\!\!\!+\!\!\sigma_w^2\|\!{\boldsymbol a}_m\!\|^2}.
\end{equation}
The impact of receive antennas on the DG performance under imperfect CSI is characterized in the following lemma.
\begin{lemma}[Receive-Antenna Scaling under Imperfect CSI]
\label{lemma:receive_antenna_scaling_imperfect_csi}
For fixed $K$, as $N_{\sf r}\to\infty$, each
device can scale its pilot power as
$p_k^{\sf p}=\bar p_k^{\sf p}/\sqrt{N_{\sf r}}$ and its transmit power as
$\tilde b_{k,m}^2=\bar b_{k,m}^2/\sqrt{N_{\sf r}}$, while almost surely maintaining the same asymptotic DG as
\begin{equation}
\label{eq:dg_simo_asymptotic_imperfect_csi}
\lim_{N_{\sf r}\to\infty}
{\bar{\mathcal{G}}}^{\boldsymbol y^{\sf f}}
=
\sum_{m=1}^{M}
\frac{\bar{\Delta}_{m}^{2}\left|\sum_{k=1}^{K}\eta_{k,m}\bar b_{k,m}\right|^{2}}
{\sum_{k=1}^{K}\eta_{k,m}^{2}|\bar b_{k,m}|^{2}{\sigma}_{m}^2
+\sigma_w^2\sum_{k=1}^{K}\eta_{k,m}},
\end{equation}
where
\begin{equation}
\label{eq:eta_imperfect_csi}
\eta_{k,m}
\triangleq
\lim_{N_{\sf r}\to\infty}\sqrt{N_{\sf r}}\varpi_k\rho_{k,m}
=
\rho_{k,m}\frac{\bar p_k^{\sf p}\varrho_k^2}{\sigma_w^2}.
\end{equation}
\end{lemma}

\begin{proof}
The key difference from the perfect CSI case is that the scaling law is governed by the estimated statistical channel $\varpi_k\rho_{k,m}$ in \eqref{eq:eta_imperfect_csi}, rather than by true statistical channel $\varrho_k$ characterized in Lemma~\ref{lemma:receive_antenna_scaling}. Specifically, each entry of the estimated channel vector $\hat{\boldsymbol h}_{k,m}$ has variance $\varpi_k\rho_{k,m}$, which scales as $1/\sqrt{N_{\sf r}}$. The remaining derivation follows the proof of Lemma~\ref{lemma:receive_antenna_scaling}. 
\end{proof}

Lemma~\ref{lemma:receive_antenna_scaling_imperfect_csi} shows that imperfect CSI weakens the transmit power benefit from receive diversity, which scales as $1/\sqrt{N_{\sf r}}$, rather than $1/N_{\sf r}$ as in the perfect CSI case.

\subsection{Extension to Nonlinear DNN Classification}
In this subsection, we leverage the insights gained from the linear classification to develop the task-aware FDM mechanism for practical nonlinear DNN classification. As analyzed previously, the task-aware design exploits the linear discriminability prior $\{{\bar{\Delta}_{m}^{2}}/{{\sigma}_{m}^2}\}$, thereby inducing non-uniform power allocation across subcarriers. However, due to the highly nonlinear nature of DNN classifiers, the linear discriminability prior $\{{\bar{\Delta}_{m}^{2}}/{{\sigma}_{m}^2}\}$ may not reflect the actual decision mechanism of the DNN. We thus resort to some empirical approaches, mainly focusing on two practical implementation issues: 
\begin{enumerate}
    \item \textit{Issue 1}: How to characterize the task prior?
    \item \textit{Issue 2}: How to derive the transceiver design?
\end{enumerate}

Generally, modern DNNs are trained through the backpropagation algorithm, which computes the gradients of the loss function w.r.t. the network parameters. Beyond training, such gradient information can be exploited to interpret model decisions. For example, well-known Grad-CAM uses the gradients flowing to measure the importance of each neuron for decision~\cite{selvaraju2017grad}. This motivates us to study \textit{Issue 1} through gradient-based prior. Formally, let the weight of the last convolutional layer of device-side submodel be denoted by $\mathbf{W} \in \mathbb{R}^{C_{\text{out}} \times C_{\text{in}} \times K_h \times K_w}$, where $C_{\text{out}}$ and $C_{\text{in}}$ denote the number of output and input channels \footnote{Here, channel refers to a dimension of feature maps rather than wireless channel in the communication system.}, and $K_h$ and $K_w$ denote the convolutional kernel height and width. Each device shares a common device-side subnetwork. The output  generates a three-dimensional feature map, denoted as $\mathbf{X} \in \mathbb{R}^{C_{\text{out}}\times L_h \times L_w}$, where $L_h$ and $L_w$ denote the height and width of each feature map.
Hence, channel-wise importance is adopted instead of element-wise importance \cite{yang2024swinjscc}. To quantify channel-wise importance, the score $g_m$ aggregates the gradients of all parameters within this specific channel-$m$, formulated as
\begin{equation}
    g_m
    =\sum_{c=1}^{C_{\mathrm{in}}}
    \sum_{u=1}^{K_h}
    \sum_{v=1}^{K_w}
    \left(\frac{\partial \mathcal{L}_{\sf DNN}}
        {\partial {W}_{m,c,u,v}}
        {W}_{m,c,u,v}
    \right)^2.
\end{equation}
 where $W_{m,c,u,v}$ denotes the $(m,c,u,v)$-th element of the weight $\mathbf{W}$ and $\mathcal{L}_{\sf DNN}$ denotes the training loss of DNN. Then, $g_m$ serves as a task-aware prior that quantifies the importance of the $m$-th feature channel.
\begin{figure}[!h]
    \centering
    \includegraphics[width=0.5\textwidth]{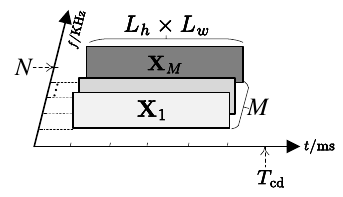}
\caption{Task-aware FDM mechanism for DNN.}
    \label{6_Fig6_FDM_channel_commom}
\end{figure}

Next, we address \textit{Issue 2}. Given the high nonlinearity of DNN, task-aware FDM for the DNN  classification cannot be directly designed through a mathematical optimization approach using statistical priors, i.e.,  $\{\bar{\Delta}_{m}^{2}, {\sigma}_{m}^2\}$ in $\mathrm{(P4)}$. Nevertheless, the insights obtained from linear classification can guide the transceiver design. Specifically, inspired by the transmitter structure in \eqref{eq:optimal_b_k_m_dg}, which weights channel inversion with the linear discriminability prior $\{{\bar{\Delta}_{m}^{2}}/{{\sigma}_{m}^2}\}$, we propose an empirically parameterized design. We first define the channel power gain as $ p_{k,m} = \tilde h_{k,m}^2$ and the active subcarrier set as $\mathcal A_k=\{m: p_{k,m}\ge p_{\sf th}\}$, where $p_{\sf th}$ denotes the channel threshold introduced to avoid excessive compensation for deeply faded subcarriers. Then, the per-subcarrier transmit power here is parameterized as
\begin{equation}
  \label{CI_PC}
  \tilde b_{k,m}^2(\beta,\gamma)=
  \begin{cases}
\frac{{P_k}}{\nu_{k,m}^2}\times\frac{g_m^\beta
p^{-\gamma}_{k,m}
}{C_b},&p_{k,m}\geq p_{\sf th}, \\
    0,& p_{k,m}< p_{\sf th},
  \end{cases}
\end{equation}
where the parameter $\beta$ controls the emphasis on task prior $g_m$, the parameter $\gamma$ determines the degree of channel inversion, and the normalization factor $C_b=\sum_{i\in\mathcal A_k}
g_i^\beta p_{k,i}^{-\gamma}$ ensures that the power constraint is satisfied. Both parameters $\beta$ and $\gamma$ are selected via grid search on the validation dataset.  Specifically, for each candidate pair $(\beta,\gamma)$, the validation loss $\mathcal{L}_{\sf val}$ is then evaluated as
\begin{equation*}
\mathcal{L}_{\sf val}(\beta,\gamma)
=\mathbb{E}_{\boldsymbol{h}_k}
\left[\mathcal{L}_{\sf DNN}\!\left(
\tilde b_{k,m}(\beta,\gamma);\boldsymbol{h}_k\right)
\right].
\end{equation*}
The optimal parameters are then determined by
\begin{equation*}
(\beta^\star,\gamma^\star)
=\underset{\beta\in\mathcal P,\,
\gamma\in\mathcal I}
{\operatorname{arg\,min}}
\;\mathcal{L}_{\sf{val}}(\beta,\gamma),
\end{equation*}
where $\mathcal{P}$ and $\mathcal{I}$ denote the searching sets for $\beta$ and $\gamma$, respectively.
The computational complexity of this method is detailed below. Let $N_{\sf val}$ denote the number of validation samples, $N_{\sf h}$ the number of channel realizations, and $\mathcal{C}_{\sf DNN}$ the computational cost of one DNN forward. For each candidate pair $(\beta,\gamma)$, computing the transmit coefficients requires $\mathcal{O}(KM)$ operations, while evaluating the validation loss requires
$\mathcal{O}\!\left(N_{\sf val}N_{\sf h}\mathcal{C}_{\sf DNN}\right)$ operations. Therefore, the overall complexity of the grid search is $\mathcal{O}\!\left(
|\mathcal{P}||\mathcal{I}|N_{\sf val}N_{\sf h}
\left(\mathcal{C}_{\sf DNN}+KM\right)
\right).$

\section{Numerical Results} \label{sec:results}


\subsection{Simulation Settings}
  We consider two representative cases: linear classification under a fitted GM model and DNN classification on the CIFAR dataset, detailed below. 
 \begin{enumerate}
     \item \emph{Linear Classification on Fitted GM}: A practical human posture classification task with $K=10$ candidate devices and $M=32$ feature dimensions is considered, where each sensing sample consists of a time–frequency spectrogram derived from FMCW signals \cite{Li2021SPAWC}. The devices are deployed at distinct spatial locations to capture different views of the target given the posture class. Each device employs \emph{principal component analysis} (PCA) for feature extraction, while the AP uses a linear classifier for final classification. The resulting feature distribution is fitted by the GM model. The statistical priors of the extracted features, i.e., $\{\boldsymbol{\mu}_\ell, \boldsymbol{\Sigma}\}$ in \eqref{Eq:class_distribution} are estimated by the noise-free features extracted from training dataset.
     \item \emph{DNN Classification on CIFAR}: Considering a DNN classification task with $K=10$ candidate devices and $M=C_{\text{out}}=64$ feature channels. The height and width of each feature map are both set to $4$, i.e., $L_h=L_w=4$. A multi-device dataset is constructed from CIFAR-10 by applying transformations, including translations, rotations, scaling, and horizontal flipping \cite{shao2022task}. The device-side employs the well-known ResNet with two standard residual blocks, while the AP-side adopts the CNN comprising two convolutional blocks followed by an MLP for final classification.
 \end{enumerate}
The entries of channel vector $\boldsymbol{h}_k$ are independently drawn from $\mathcal{CN}(0,1)$ and $\sigma_w^2$ is set according to the corresponding transmit SNR. All simulation results are averaged over 1,000 independent Monte Carlo trials.

\begin{figure}[!htb]
	\centering
    	\begin{minipage}{0.3\textwidth}
		{\includegraphics[width=\textwidth]
			{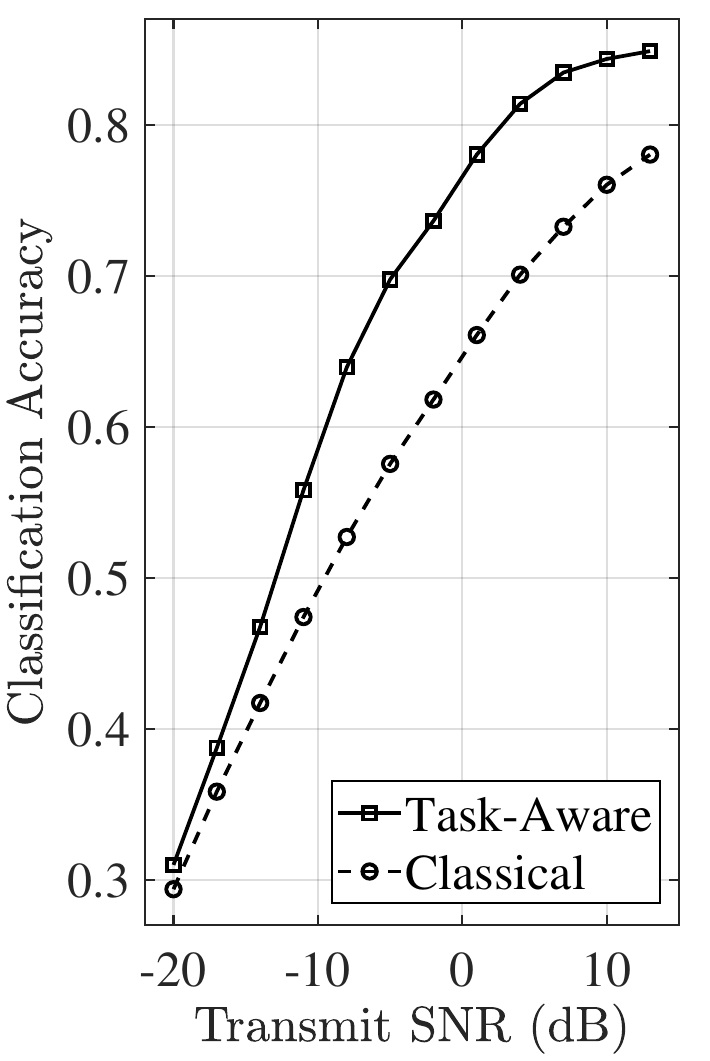}}
		\caption*{(a)}\label{Fig_MSE_vs_DG_MLP_SNRp}        
	\end{minipage}
	\begin{minipage}{0.3\textwidth}
		{\includegraphics[width=\textwidth]
			{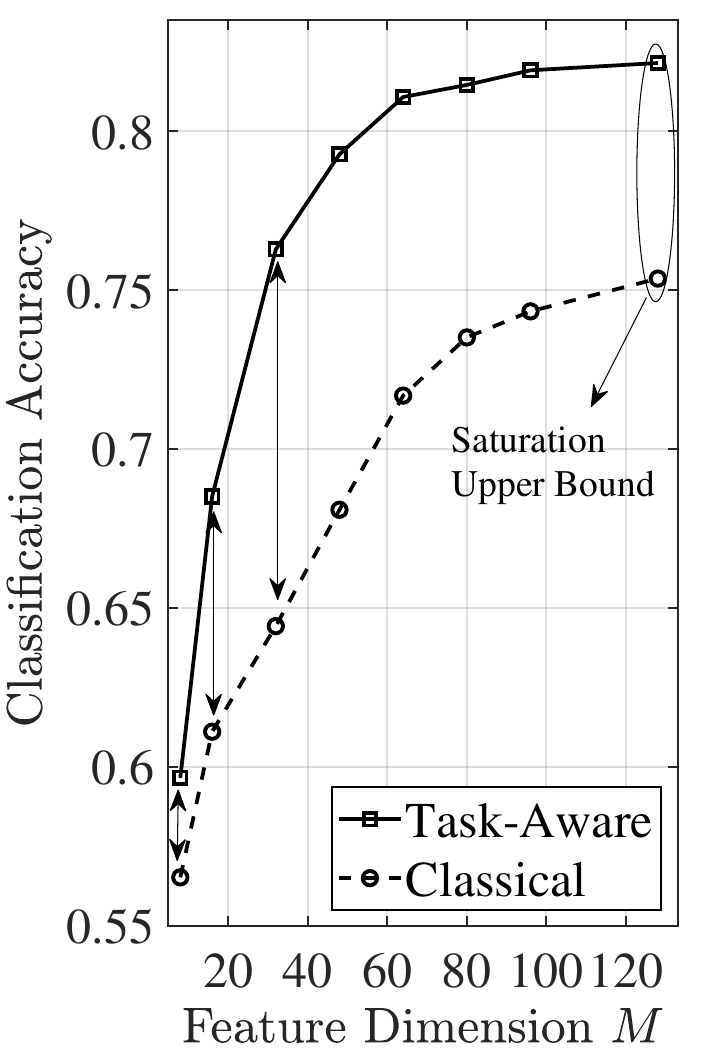}}
		\caption*{(b)}\label{Fig_MSE_vs_DG_MLP_M}        
	\end{minipage}
	\begin{minipage}{0.3\textwidth}
		{\includegraphics[width=\textwidth]
			{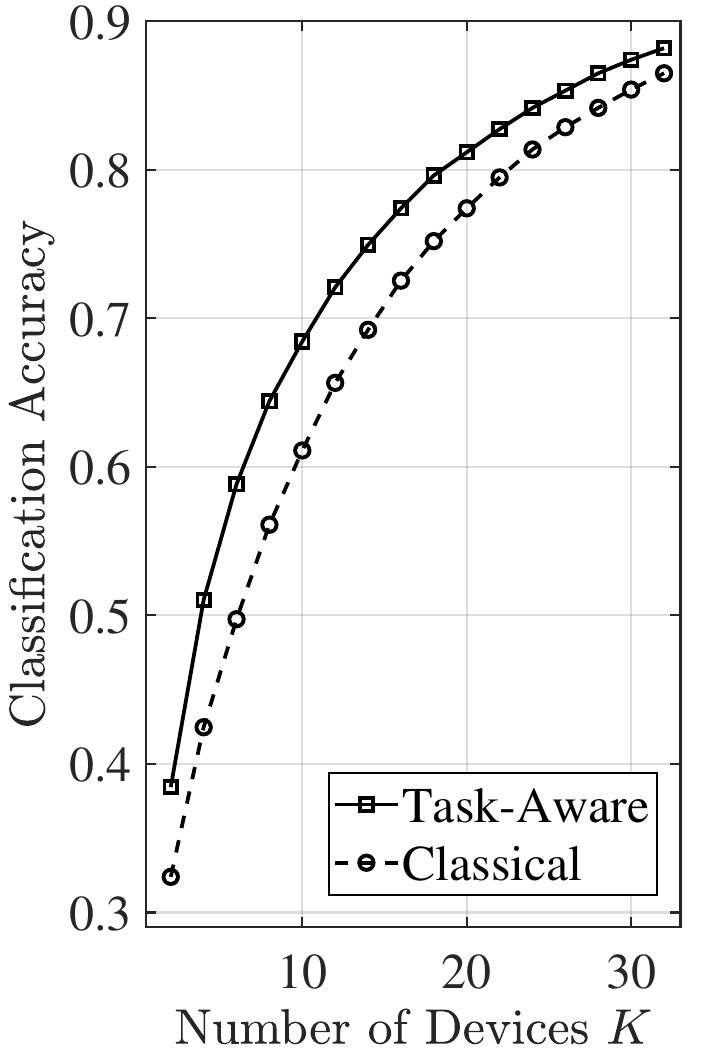}}
		\caption*{(c)}\label{Fig_MSE_vs_DG_MLP_K}
	\end{minipage}
	
	\caption{Linear classification performance comparison under varying transmit SNRs, number of feature dimension $M$, and number of devices $K$.}\label{Fig_P0_acc_KN}
\end{figure}

\subsection{Performance Analysis: Linear Classification}

Fig.~\ref{Fig_P0_acc_KN} compares the classification accuracy of the classical AirComp and task-aware designs under varying transmit SNRs, number of feature dimensions $M$, and number of devices $K$. Three key observations can be made: First, Fig.~\ref{Fig_P0_acc_KN}(a) shows that both designs suffer from severe performance degradation at extremely low SNRs. As the SNR increases, the task-aware aggregation achieves superior classification accuracy by enabling task-aware resource allocation across features, consistent with the numerical results in Fig.~\ref{Fig_P_acc_snr_FDM} and the discussion in Section~\ref{sec:Discussion_Aware}. Second, Fig.~\ref{Fig_P0_acc_KN}(b) shows that the classification accuracy improves as the feature dimension $M$ increases. The performance gap initially widens; then two curves gradually approach their respective saturation upper bounds. This is because increasing $M$ provides more task-relevant feature information, thereby enhancing the aggregation quality. Third, Fig.~\ref{Fig_P0_acc_KN}(c) shows that the classification accuracy improves as more observations are available for aggregation. However, the underlying reason is different: increasing the number of devices $K$ suppresses the effective sensing noise through aggregation, thereby yielding a more reliable feature.

\begin{figure}[!htbp]
	\centering
	\begin{minipage}{0.3\textwidth}
		{\includegraphics[width=\textwidth]
			{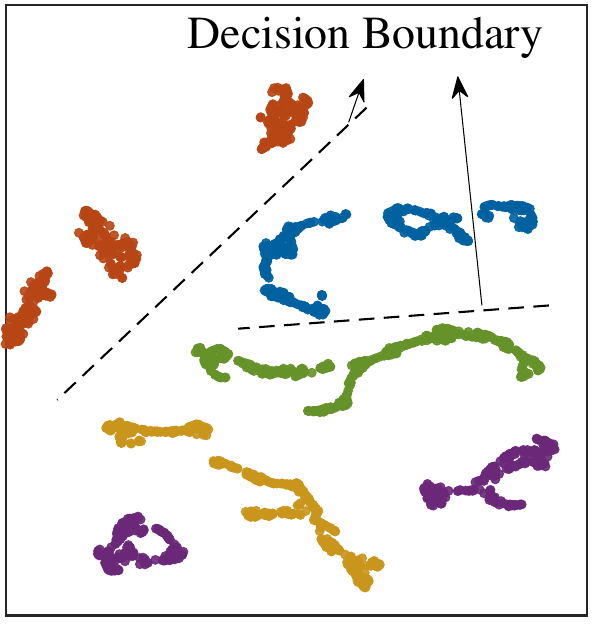}}
		\caption*{(a)}\label{Fig_P4tsne_snr_10_X}
	\end{minipage}
	\begin{minipage}{0.3\textwidth}
		{\includegraphics[width=\textwidth]
			{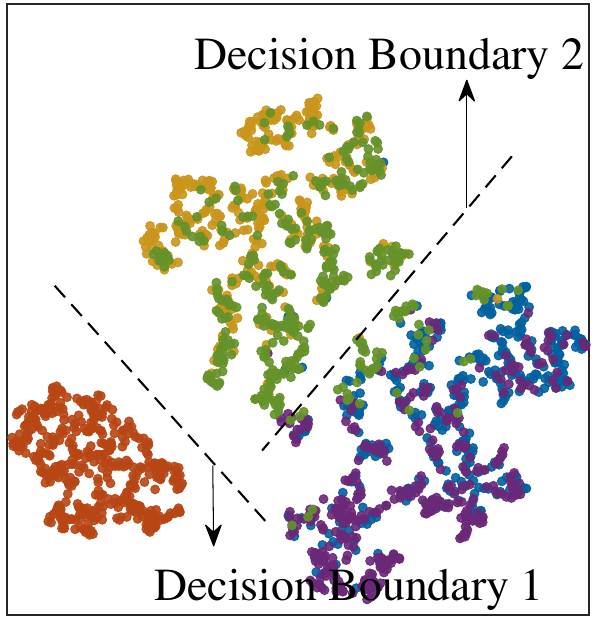}}
		\caption*{(b)}\label{Fig_P4tsne_snr_10_Xhat_dg}
	\end{minipage}
	\begin{minipage}{0.3\textwidth}
		{\includegraphics[width=\textwidth]
			{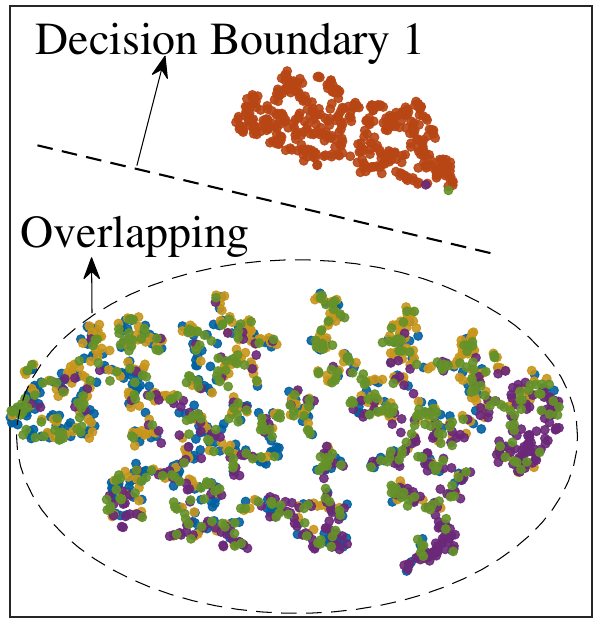}}
		\caption*{(c)}\label{Fig_P4tsne_snr_10_Xhat_mse}
	\end{minipage}
	\caption{The feature visualization: (a) raw features, (b) task-aware, (c) classical AirComp.}
	\label{Fig_results_tsne}
\end{figure}

Fig.~\ref{Fig_results_tsne} visualizes the feature distributions. Specifically, t-SNE is employed to project the features from the original high-dimensional space onto a two-dimensional plane \cite{maaten2008visualizing}.  Fig.~\ref{Fig_results_tsne} shows that the task-aware aggregation yields clearer inter-class separation. For example, the task-aware design reveals two distinct decision boundaries between clusters, indicating richer separability for the classification task, whereas the classical aggregation exhibits essentially a single dominant decision boundary with noticeable cluster overlapping.

\begin{figure}[!htb]
	\centering
    	\begin{minipage}{0.3\textwidth}
		{\includegraphics[width=\textwidth]
			{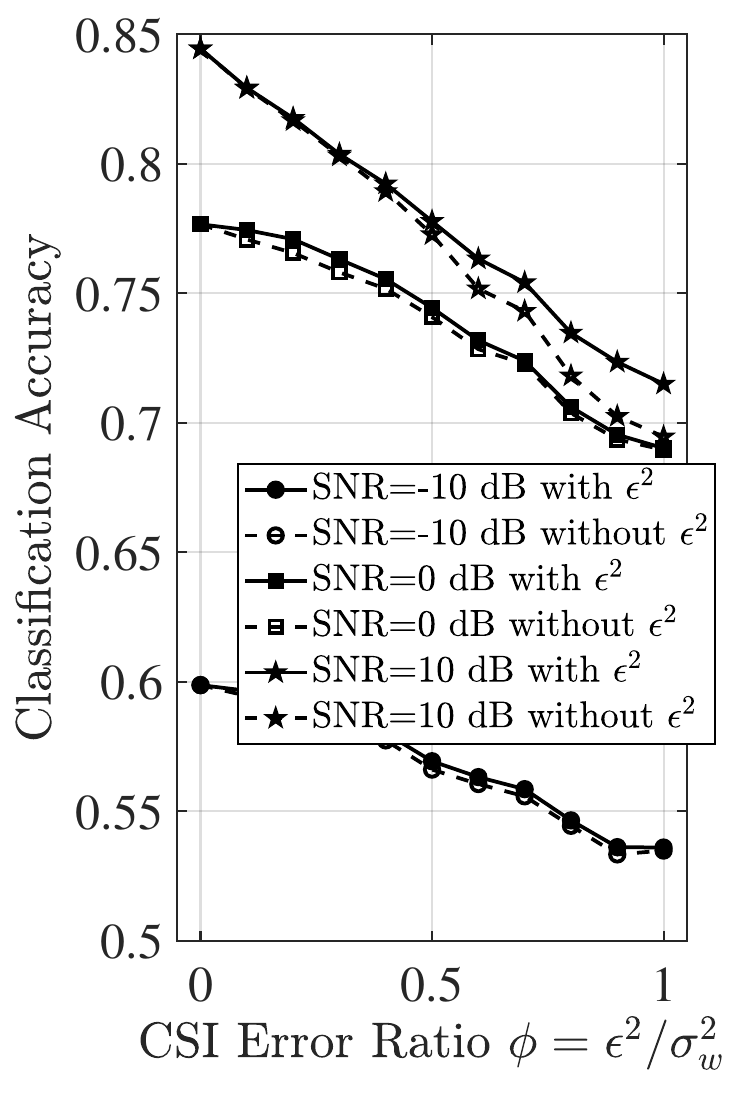}}
		\caption*{(a)}\label{MSE_vs_DG_MLP_imperfect_CSI}        
	\end{minipage}
	\begin{minipage}{0.3\textwidth}
		{\includegraphics[width=\textwidth]
			{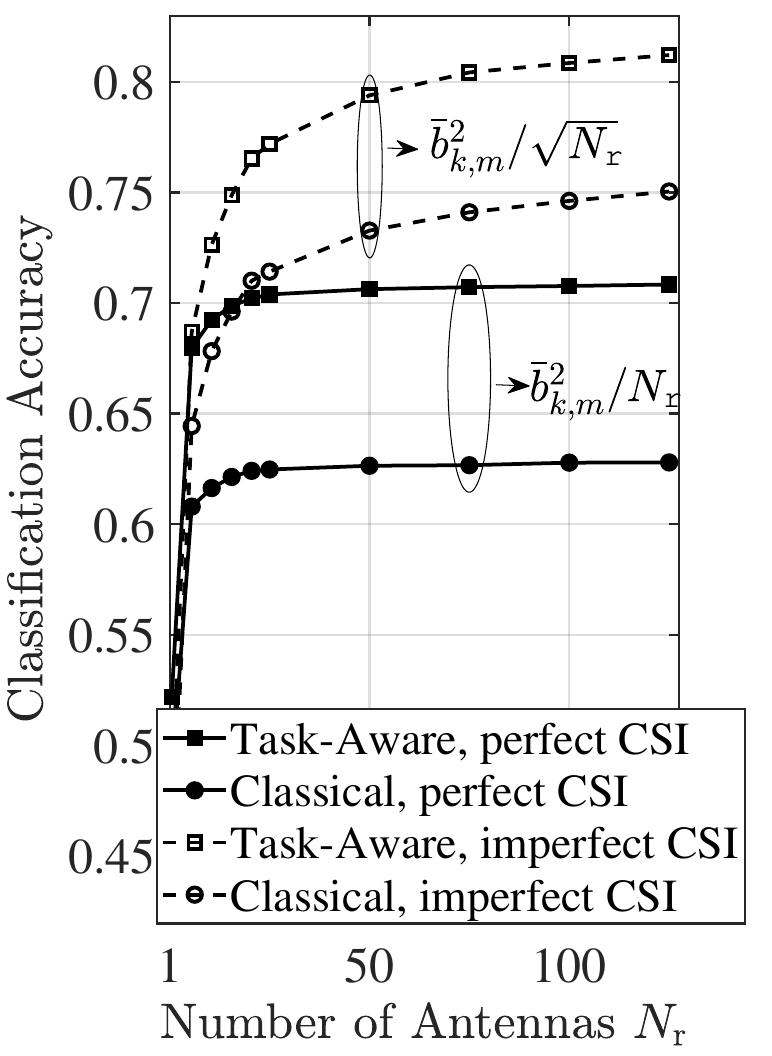}}
		\caption*{(b)}\label{Fig_MSE_vs_DG_MLP_M}        
	\end{minipage}
	
	\caption{Classification performance comparison under different CSI error ratios and varying numbers of receive antennas $N_{\sf r}$.}\label{MSE_vs_DG_MLP_CSI_antenna}
\end{figure}

In Fig.~\ref{MSE_vs_DG_MLP_CSI_antenna}(a), the impact of imperfect CSI on classification accuracy with different transmit SNRs is given. As discussed after Proposition~\ref{proposition:per_subcarrier_csi_error_model}, the CSI error $\epsilon_{k,m}^{2}$ is determined by three factors. Generally, the ML estimator yields $\epsilon_{k,m}^{2}=\sigma_{w}^{2}$, whereas the LMMSE estimator in~\eqref{eq:received_csi} satisfies $\epsilon_{k,m}^{2}\leq \sigma_{w}^{2}$. We then define the CSI error ratio as $\phi \triangleq \frac{\epsilon^{2}}{\sigma_{w}^{2}}$, where $\epsilon^{2}=\epsilon_{k,m}^{2}$. Specifically, $\phi=1$ corresponds to ML estimation and $0\leq\phi<1$ holds for LMMSE estimation~\cite{xie2023optimal}. A key observation from Fig.~\ref{MSE_vs_DG_MLP_CSI_antenna}(a) is that the classification accuracy becomes increasingly sensitive to $\phi$ in the high-SNR regime. For example, the performance gap between imperfect and perfect CSI widens as $\phi$ increases when SNR equals $10$~dB. This is because, in the absence of regularization, CSI errors are amplified as the transmit power increases.
\begin{figure}[!htb]
	\centering
    	\begin{minipage}{0.3\textwidth}
		{\includegraphics[width=\textwidth]
			{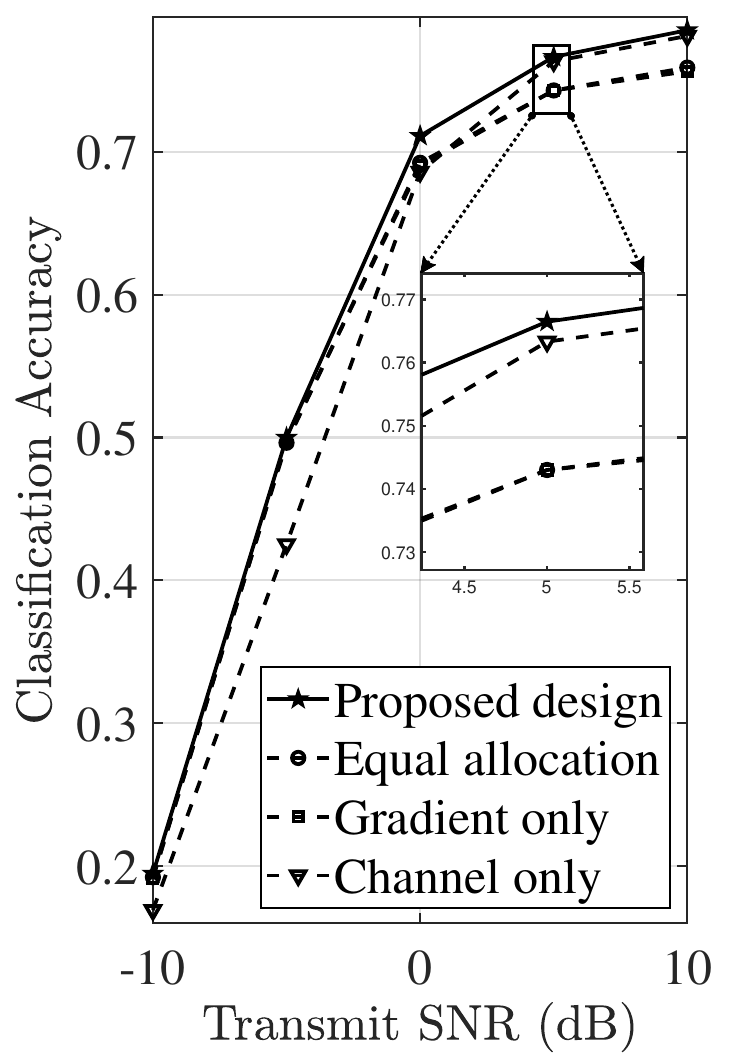}}
		\caption*{(a)}\label{Fig_MSE_vs_DG_MLP_SNRp}        
	\end{minipage}
	\begin{minipage}{0.3\textwidth}
		{\includegraphics[width=\textwidth]
			{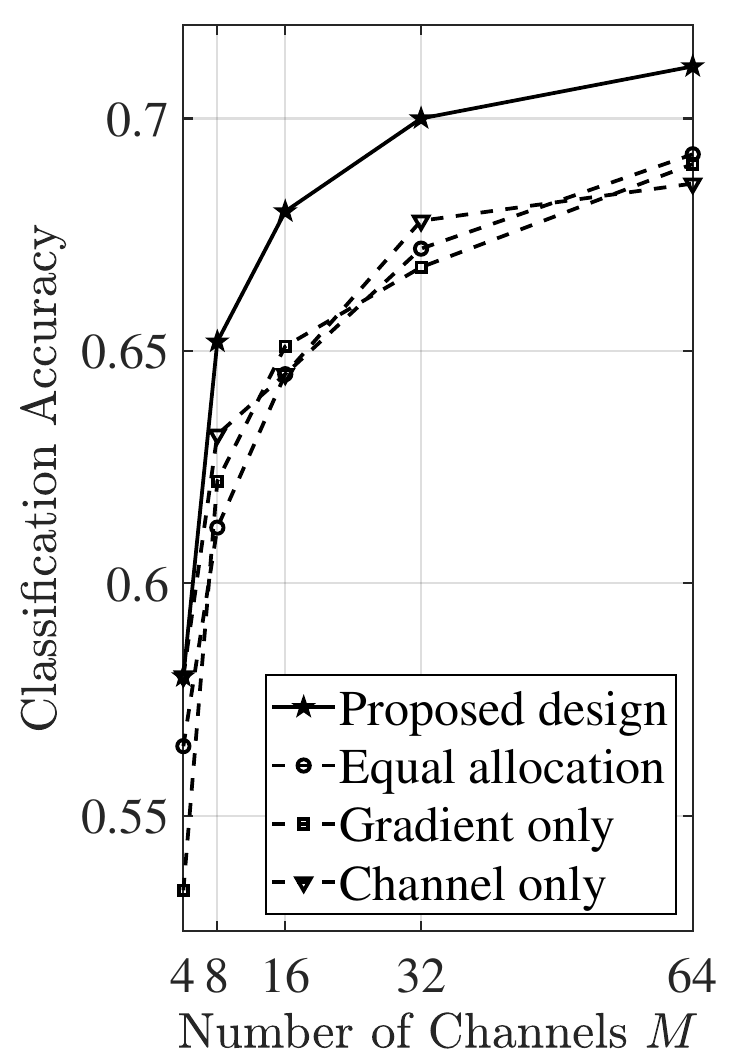}}
		\caption*{(b)}\label{Fig_MSE_vs_DG_MLP_M}        
	\end{minipage}
	\begin{minipage}{0.3\textwidth}
		{\includegraphics[width=\textwidth]
			{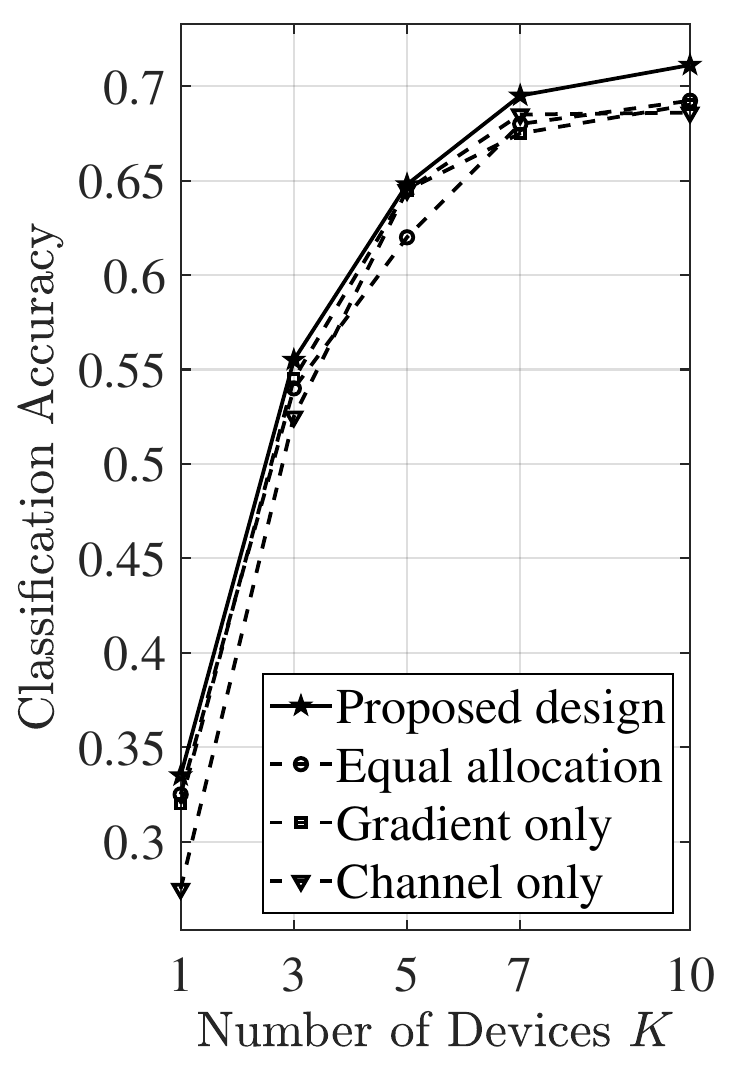}}
		\caption*{(c)}\label{Fig_MSE_vs_DG_MLP_K}
	\end{minipage}
	
	\caption{DNN classification performance comparison under varying transmit SNRs, number of feature channels $M$, and number of devices $K$.}\label{Fig_P1_acc_KN}
\end{figure}
\begin{figure}[!h]
    \centering
    \includegraphics[width=0.4\textwidth]{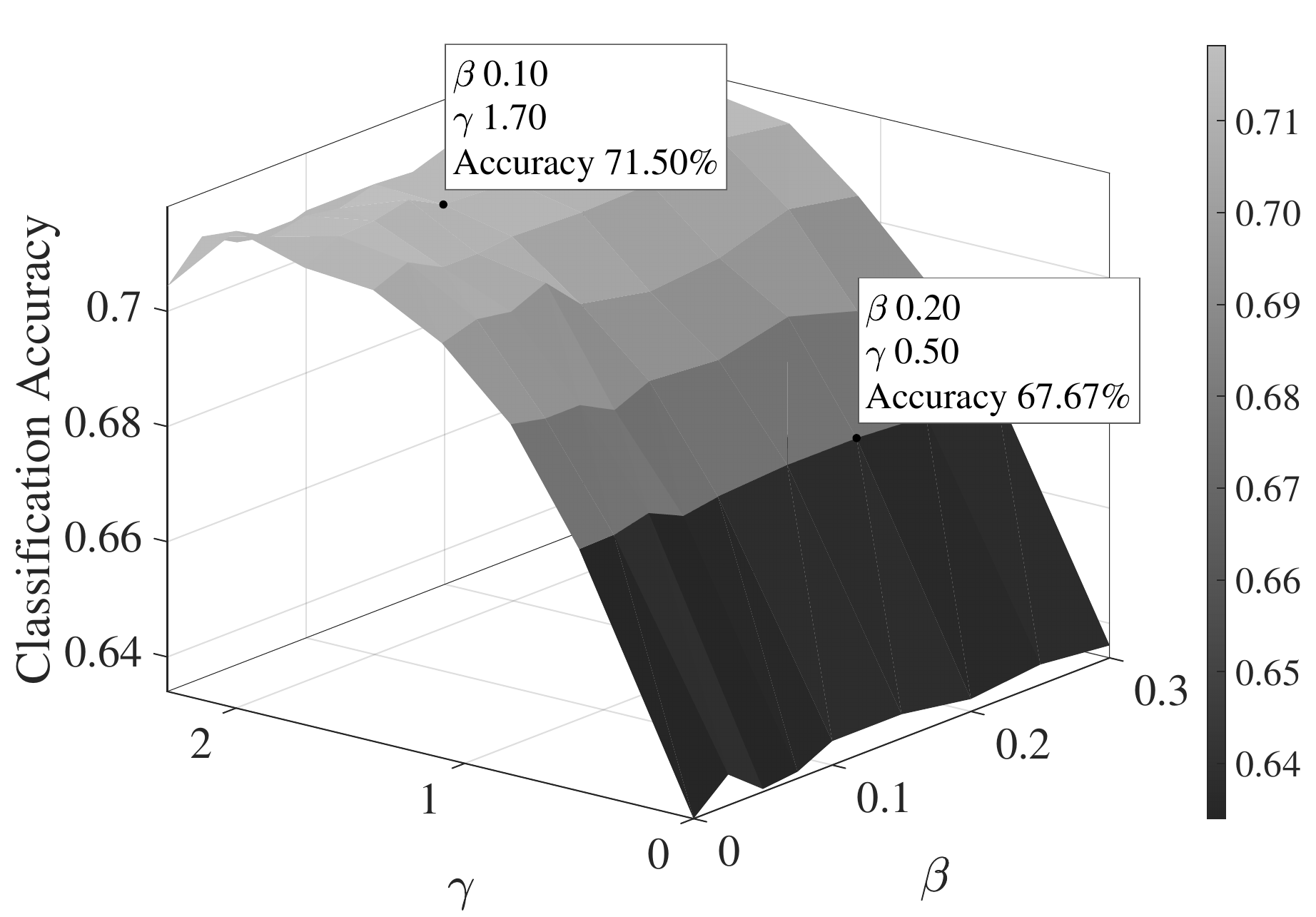}
\caption{Results of grid search on the validation dataset.}
    \label{grid_serach_3D}
\end{figure}
Fig.~\ref{MSE_vs_DG_MLP_CSI_antenna}(b) validates the receive-antenna scaling law of the classification performance. Here, as the number of receive antennas increases, the transmit scaling follows $\tilde{b}_{k,m}^{2}=\bar{b}_{k,m}^{2}/N_{\sf r}$ under perfect CSI and $\tilde{b}_{k,m}^{2}=\bar{b}_{k,m}^{2}/\sqrt{N_{\sf r}}$ under imperfect CSI. It can be observed that the performance of all schemes converges to a constant, consistent with Lemma~\ref{lemma:receive_antenna_scaling} and Lemma~\ref{lemma:receive_antenna_scaling_imperfect_csi}. An interesting observation is that the imperfect CSI case outperforms the perfect CSI case. This is because their transmit scaling laws are different. Specifically, $\bar{b}_{k,m}^{2}/\sqrt{N_{\sf r}}$ preserves more effective transmit power than $\bar{b}_{k,m}^{2}/N_{\sf r}$, thereby compensating for the degradation caused by CSI errors.

\subsection{Performance Analysis: DNN Classification}
Fig.~\ref{Fig_P1_acc_KN} compares the proposed task-aware FDM mechanism with three baselines: \emph{Equal Allocation}, which assigns equal power to all subcarriers; \emph{Gradient Only}, which considers only the gradient prior in \eqref{CI_PC}; and \emph{Channel Only}, which considers only channel information in \eqref{CI_PC}. The observations are consistent with those for linear classification in Fig.~\ref{Fig_P0_acc_KN}. Specifically, the proposed mechanism outperforms all baselines across varying transmit SNRs, feature channels $M$, and numbers of devices $K$. This is because the proposed empirically parameterized design jointly captures the effects of channel conditions and task prior. For example, $0<\gamma<1$ captures channel inversion while $\beta>0$ introduces task prior. The parameters of the proposed design are selected via grid search, whose results are visualized in Fig.~\ref{grid_serach_3D}. As shown, different combinations of $\beta$ and $\gamma$ yield different classification accuracies.

\section{Conclusion} \label{sec:conclusion}
In this paper, we have investigated how to maximize classification performance in the AirComp-enabled ISCC system. Guided by the theoretical analysis, a task-aware FDM transmission mechanism is developed, which realizes task-aware aggregation for linear classification. Analytical and numerical comparisons with the classical aggregation reveal that the developed FDM mechanism essentially performs non-uniform power allocation across subcarriers, thereby striking a tradeoff between the inter-class separation $\bar{\Delta}_{m}^2$ and the intra-class variance $\sigma_{m}^2$. This mechanism is further generalized to the imperfect CSI setting, where the prior-guided structure is maintained by treating CSI error as a  regularization term. Moreover, we further show that each device can scale down its transmit power as $1/N_{\sf r}$ and $1/\sqrt{N_{\sf r}}$ under perfect and imperfect CSI, respectively, without degrading the DG performance. Since the DNN classification is analytically intractable, an empirically parameterized design is thus proposed, which is guided by the insights obtained from linear classification and combines channel inversion with gradient-prior weighting. Future research may investigate the robustness of the proposed framework under multimodal sources, as well as dynamic subcarrier selection.

\appendix

\subsection{Proof of Proposition~\ref{proposition:dg_optimal_TDM}}\label{appendix:Proof of proposition1}
Let $c_{k,m} \triangleq \tilde{h}_k\tilde{b}_{k,m}$ and
$u_{k,m} \triangleq \tilde{h}_k\sqrt{P_k}/\nu_{k,m}$. Then, $(\mathrm{P}2)$ can be equivalently reformulated as
\begin{align*}
(\mathrm{P}2.1)\quad\quad\quad
&\max_{\{c_{k,m}\}}\quad
\varphi(c_{k,m})
=\frac{\bar{\Delta}_{m}^2\left(\sum_{k=1}^{K} c_{k,m}\right)^2}{
\sum_{k=1}^{K} {\sigma}_{m}^2 c_{k,m}^2+\sigma_w^2}\quad\quad\\
&\;\;\;\text{s.t.}\quad
0\le c_{k,m}\le u_{k,m},\ \forall k.
\end{align*}
Let $c_{\sf sum}\triangleq \sum_{k=1}^{K}c_{k,m}$ denote the effective aggregate gain. For any
$c_{\sf sum}$, maximizing $\varphi(c_{k,m})$ is equivalent to minimizing
the denominator term $\sum_{k=1}^{K}{\sigma}_{m}^2c_{k,m}^2$ in $(\mathrm{P}2.1)$.
Therefore, $(\mathrm{P}2.1)$ can be reformulated as
\begin{align*}
\mathrm{(P2.2)}
\min_{\{c_{k,m}\}}\!
\sum_{k=1}^{K}{\sigma}_{m}^2 c_{k,m}^2 
\;\;\text{s.t.}  \sum_{k=1}^{K} c_{k,m}=c_{\sf sum}, c_{k,m}\le u_{k,m},
\end{align*} 
which is a quadratic programming problem and its Lagrangian is given by 
\begin{align} \mathcal{L}_{\mathrm{P}2.2}
=&\sum_{k=1}^{K}{\sigma}_{m}^2 c_{k,m}^2
-\lambda\left(\sum_{k=1}^{K}c_{k,m}-c_{\sf sum}\right)+\sum_{k=1}^{K}\alpha_k(c_{k,m}-u_{k,m})
-\sum_{k=1}^{K}\xi_k c_{k,m},\label{eq:LP13}
\end{align}
where $\lambda$, $\alpha_k\ge0$, and $\xi_k\ge0$ denote the Lagrange
multipliers. According to the Karush--Kuhn--Tucker conditions
\cite{boyd2004convex}, the optimal solution satisfies
\begin{equation*}
2{\sigma}_{m}^2c_{k,m}-\lambda+\alpha_k-\xi_k=0,\quad \forall k.
\end{equation*}
Hence, the optimal solution exhibits a threshold-based
structure as
\begin{equation}\label{eq:weighted_capped_structure}
c_{k,m}^\star
=\min\left\{
u_{k,m},
\frac{\tau}{{\sigma}_{m}^2}
\right\},
\qquad
\tau\triangleq\frac{\lambda}{2}\ge0.
\end{equation}
To determine the threshold $\tau$, we define $\theta_k\triangleq {\sigma}_{m}^2u_k$. Without loss of generality, let the devices be ordered as
$\theta_1\le \theta_2\le \cdots\le \theta_K.$ This ordering implies that the devices with smaller $\theta_k$ saturate first. Suppose that the first $j$ devices are saturated, whereas the remaining devices follow the threshold rule, i.e.,
\begin{equation}\label{eq:weighted_segment_structure}
c_{k,m}
=
\begin{cases}
u_{k,m}, & 1\le k\le j,\\[6pt]
\dfrac{\tau}{{\sigma}_{m}^2}, & j<k\le K.
\end{cases}
\end{equation}
Substituting \eqref{eq:weighted_segment_structure} into $\varphi(c_{k,m})$ in $(\mathrm{P}2.1)$ yields 
\begin{equation}\label{eq:varphi}
    \varphi_j(\tau)
=\frac{\bar{\Delta}_{m}^2\left(\sum_{k=1}^{j}u_{k,m}+\tau\sum_{k=j+1}^{K}\frac{1}{{\sigma}_{m}^2}\right)^2}{\sum_{k=1}^{j}{\sigma}_{m}^2u_{k,m}^2
+\tau^2\sum_{k=j+1}^{K}\frac{1}{{\sigma}_{m}^2}+\sigma_w^2.} 
\end{equation}
By checking the first-order derivative of \eqref{eq:varphi}, the stationary point
within the $j$-th segment is given by 
\begin{equation*}
\tau_j
=\frac{\sigma_w^2+\sum_{k=1}^{j}{\sigma}_{m}^2u_{k,m}^2}{\sum_{k=1}^{j}u_{k,m}},
\qquad j\ge1.
\end{equation*}
If $\theta_j\le \tau_j < \theta_{j+1}$, $\tau_j$ is feasible for this
segment and serves as the maximizer. Otherwise, the segment-wise
optimum occurs at the boundary. By comparing all feasible segments, the global
threshold $\tau^\star$ can be obtained. Accordingly, the optimal effective
receive amplitudes $c_{k,m}^\star$ are determined and the optimal transmit amplitudes are
given by \eqref{eq:p3_solution}.

\subsection{Proof of Proposition~\ref{proposition:decision_optimal_fdm}}
\label{appendix:p2_lagrange_duality}
The Lagrangian of $(\mathrm{P}4)$ and its dual function are respectively given by
\begin{equation*}
\mathcal{L}_{\mathrm{P}4}
=\sum_{m=1}^{M}\left[\Phi_m(\{\tilde{b}_{k,m}\}_{k=1}^{K};\boldsymbol\lambda)\right]
+\sum_{k=1}^{K}\lambda_k P_k,
\end{equation*}
\begin{equation}\label{eq:dualP6_dg}
g(\boldsymbol\lambda)
=\sum_{m=1}^{M}\underbrace{\sup_{\{\tilde{b}_{k,m}\}_{k=1}^{K}}\Phi_m(\{\tilde{b}_{k,m}\}_{k=1}^{K};\boldsymbol\lambda)}_{\psi_m(\boldsymbol\lambda)}
+\sum_{k=1}^{K}\lambda_k P_k,
\end{equation}
where
\begin{equation*}
\Phi_m(\{\tilde{b}_{k,m}\}_{k=1}^{K};\boldsymbol\lambda)\!\!
\triangleq\!
\frac{\bar{\Delta}_{m}^2\left(\sum_{k=1}^{K}\tilde{h}_{k,m}\tilde{b}_{k,m}\right)^2}
{\sum_{k=1}^{K}\tilde{h}_{k,m}^2\tilde{b}_{k,m}^2{\sigma}_{m}^2+\sigma_w^2}
\!-\!\!\!\sum_{k=1}^{K}\lambda_k\nu_{k,m}^2\tilde{b}_{k,m}^2,
\end{equation*}
and $\boldsymbol{\lambda}=[\lambda_1,\ldots,\lambda_K]^{\mathsf T}\succeq\mathbf 0$ collects the dual variables associated with the individual power constraints of the devices. Since $(\mathrm P4)$ is a maximization problem, its dual problem is to minimize $g(\boldsymbol\lambda)$ over $\boldsymbol\lambda\succeq\mathbf 0$. For each subcarrier $m$, we define an auxiliary variable as \begin{equation}\label{eq:z_m_definition_dg}
z_m\triangleq
\frac{\sum_{k=1}^{K}\tilde{h}_{k,m}\tilde{b}_{k,m}}
{\sum_{k=1}^{K}\tilde{h}_{k,m}^2\tilde{b}_{k,m}^2{\sigma}_{m}^2+\sigma_w^2}.
\end{equation}
Taking the derivative of $\Phi_m(\{\tilde{b}_{k,m}\}_{k=1}^{K};\boldsymbol\lambda)$ w.r.t. $\tilde{b}_{k,m}$ yields the stationary point
\begin{equation}\label{eq:optimal_b_k_m_dg}
\tilde{b}_{k,m}^{\star}(\lambda_k,z_m)
=
\frac{\bar{\Delta}_{m}^2\tilde{h}_{k,m}z_m}
{\lambda_k\nu_{k,m}^2+\bar{\Delta}_{m}^2{\sigma}_{m}^2\tilde{h}_{k,m}^2z_m^2}.
\end{equation}
Substituting \eqref{eq:optimal_b_k_m_dg} into \eqref{eq:z_m_definition_dg} gives
\begin{equation}\label{eq:consistency_condition_dg}
\sum_{k=1}^{K}
\frac{\lambda_k\tilde{h}_{k,m}^2\nu_{k,m}^2}
{\left(\lambda_k\nu_{k,m}^2+\bar{\Delta}_{m}^2\tilde{h}_{k,m}^2z_m^2{\sigma}_{m}^2\right)^2}
=
\frac{\sigma_w^2}{\bar{\Delta}_{m}^2}.
\end{equation}
The left-hand side of \eqref{eq:consistency_condition_dg} is monotonically decreasing w.r.t. $z_m^2$ when $\lambda_k>0$. Therefore, if $\sum_{k=1}^{K}\frac{\tilde{h}_{k,m}^2}{\lambda_k\nu_{k,m}^2}>\frac{\sigma_w^2}{\bar{\Delta}_{m}^2},$ \eqref{eq:consistency_condition_dg} admits a non-zero solution. Substituting $z_m^\star(\boldsymbol\lambda)$ into \eqref{eq:optimal_b_k_m_dg} gives the solution of the $m$-th Lagrangian subproblem. Then, the dual function is
\begin{equation}\label{eq:g_lambda_dg}
g(\boldsymbol\lambda)
=
\sum_{m=1}^{M}\Phi_m\left(\{\tilde{b}_{k,m}^{\star}(\lambda_k,z_m^\star)\}_{k=1}^{K};\boldsymbol\lambda\right)
+\sum_{k=1}^{K}\lambda_kP_k.
\end{equation}
Based on \eqref{eq:optimal_b_k_m_dg}, \eqref{eq:consistency_condition_dg}, and power constraint, the dual problem can be solved using subgradient-based methods. A subgradient of $g(\boldsymbol\lambda)$ w.r.t. $\lambda_k$ is given by
$s_k(\boldsymbol\lambda)=P_k-\sum_{m=1}^{M}\nu_{k,m}^2(\tilde{b}_{k,m}^{\star})^2.$
At the optimum, the complementary slackness condition satisfies $\lambda_k^\star\left(\sum_{m=1}^{M}\nu_{k,m}^2|\tilde{b}_{k,m}^{\star}|^2-P_k\right)=0, \forall k.$ Then, the outer-layer variables $\{\lambda_k\}$ are updated to satisfy the power constraint based on \eqref{eq:consistency_condition_dg}, while the inner layer solves for the optimal $z_m^\star$ on each subcarrier according to \eqref{eq:consistency_condition_dg}. The overall procedure is summarized in Algorithm~\ref{alg:dual_P2_dg}. The convergence and complexity of the proposed algorithm are discussed below. 

\textit{Convergence}: It can be verified that \eqref{eq:consistency_condition_dg} admits at most one positive solution. For the outer layer update, the projected subgradient method converges to a neighborhood of the optimal dual value, where the neighborhood size decreases with the step size. The objective in $(\mathrm{P}4)$ involves a fractional programming and is thus non-convex, the dual optimum provides an upper bound on the primal optimum \cite{boyd2004convex}. When the time-sharing condition holds, the primal variables recovered from \eqref{eq:optimal_b_k_m_dg} are globally optimal \cite{yu2006dual}. 

\textit{Complexity}: For a given $\boldsymbol{\lambda}$, the left hand of \eqref{eq:consistency_condition_dg} can be evaluated with complexity $\mathcal{O}(K)$. The optimal $z_m^\star$ can be
obtained by bisection. Let $I_z$ denote the number of bisection iterations. Then, the complexity of solving $\{z_m^\star\}_{m=1}^{M}$ is $\mathcal{O}(MKI_z)$. After obtaining $\{z_m^\star\}_{m=1}^{M}$, 
$\{\tilde{b}_{k,m}^{\star}\}$ are updated by
\eqref{eq:optimal_b_k_m_dg}, which requires $\mathcal{O}(MK)$ operations.
The computation of the power violations for updating the dual variables requires $\mathcal{O}(MK)$ operations. Therefore, the computational
complexity per outer dual iteration is $\mathcal{O}\!\left(MKI_z+MK\right)
=\mathcal{O}\!\left(MK(I_z+1)\right).$
Let $I_{\lambda}$ denote the number of outer iterations required for dual
convergence; the overall complexity is $\mathcal{O}\!\left(I_{\lambda}MK(I_z+1)\right).$

\subsection{Proof of Corollary~\ref{corollary:mse_equivalent_objective}}
\label{appendix:mse_equivalent_obj}
We first transform \eqref{Eq:AirComp_statistical_error} as follows. For fixed $\{\tilde{b}_{k,m}\}_{k=1}^{K}$, the $m$-th term of \eqref{Eq:AirComp_statistical_error} is quadratic w.r.t. $\tilde{a}_m$, i.e.,
\begin{equation}\label{eq:explanation_a_m}
\mathcal{D}_m(\tilde{a}_m)
= \sum_{k=1}^{K}(\tilde{a}_m\tilde{h}_{k,m}\tilde{b}_{k,m}-1)^2\nu_{k,m}^2+\tilde{a}_m^2\sigma_w^2,
\end{equation}
whose minimizer is 
\begin{equation*}
\tilde{a}_m^\star(\{\tilde{b}_{k,m}\}_{k=1}^{K})
= \frac{\sum_{k=1}^{K}\tilde{h}_{k,m}\tilde{b}_{k,m}\nu_{k,m}^2}
{\sum_{k=1}^{K}\tilde{h}_{k,m}^2\tilde{b}_{k,m}^2\nu_{k,m}^2+\sigma_w^2}.
\end{equation*}
Substituting $\tilde{a}_m^\star$ into \eqref{eq:explanation_a_m} yields
\begin{equation*}
\mathcal{D}_m^\star(\{\tilde{b}_{k,m}\}_{k=1}^{K})\!\!
= \!\!\sum_{k=1}^{K}\nu_{k,m}^2\!\!\!
-\!\!\underbrace{\frac{\left(\sum_{k=1}^{K}\tilde{h}_{k,m}\tilde{b}_{k,m}\nu_{k,m}^2\right)^2}
{\sum_{k=1}^{K}\tilde{h}_{k,m}^2\tilde{b}_{k,m}^2\nu_{k,m}^2\!\!+\!\!\sigma_w^2}}_{\mathcal U_m}.
\end{equation*}
Since $\sum_{k=1}^{K}{\nu_{k,m}^2}$ is independent of $\{\tilde{b}_{k,m}\}_{k=1}^{K}$, minimizing $\mathcal{D}_m$ is equivalent to maximizing $\mathcal U_m$. This completes the proof.

\subsection{Proof of Lemma~\ref{lemma:receive_antenna_scaling}}
\label{appendix:receive_antenna_scaling}

Let the AP adopt the \emph{maximum-ratio combining} (MRC) receivers $\boldsymbol{a}_m
=\frac{1}{\sqrt{N_{\sf r}}}
\sum_{k=1}^{K}\boldsymbol{h}_{k,m},$ we have
\begin{equation*}
\lim_{N_{\sf r}\to\infty}\frac{1}{\sqrt{N_{\sf r}}}
\boldsymbol{a}_m^{\mathsf H}\boldsymbol{h}_{k,m}
\overset{(a)}{=}\varrho_{k},~
\lim_{N_{\sf r}\to\infty}\|\boldsymbol{a}_m\|^2
\overset{(b)}{=}
\sum_{k=1}^{K}\varrho_{k},
\end{equation*}
where (a) and (b) follow the channel
hardening and asymptotic orthogonality properties as \cite{ngo2013energy}
\begin{equation*}
\lim_{N_{\sf r}\to\infty} \frac{1}{N_{\sf r}}\boldsymbol{h}_{i,m}^{\mathsf H}\boldsymbol{h}_{k,m}=
\begin{cases}
\varrho_{k}, & i=k,\\
0, & i\neq k.
\end{cases}
\end{equation*}
Substituting these asymptotic results into \eqref{eq:dg_fdm_simo} gives \eqref{eq:dg_simo_asymptotic}. We can arrive at the same result for the \emph{zero forcing} (ZF) and \emph{minimum mean-square error} (MMSE) receivers; the details are omitted here. This completes the proof.


\subsection{Proof of Proposition~\ref{proposition:per_subcarrier_csi_error_model}}
\label{appendix:imperfect_csi_model}
The LMMSE estimate in \eqref{eq:received_csi} can be rewritten as
\begin{equation}\label{eq:lmmsecsi}
\hat{\boldsymbol h}_k^{\sf p}=\boldsymbol h_k^{\sf p}+\boldsymbol e_k^{\sf p}.
\end{equation}
where $\boldsymbol e_k^{\sf p}$ denotes the pilot estimation error; $\hat{\boldsymbol h}_k^{\sf p}$
and $\boldsymbol e_k^{\sf p}$ are independent and distributed as
\begin{align*}
\hat{\boldsymbol h}_k^{\sf p}&\sim\mathcal{CN}\left(\boldsymbol 0,\varpi_k\mathbf I_{N_{\sf p}}\right),\qquad
\varpi_k=\frac{p_k^{\sf p}\varrho_k^2}{p_k^{\sf p}\varrho_k+\sigma_w^2},\\
\boldsymbol e_k^{\sf p}&\sim\mathcal{CN}\left(\boldsymbol 0,\kappa_k\mathbf I_{N_{\sf p}}\right),\qquad
\kappa_k=\frac{\varrho_k\sigma_w^2}{p_k^{\sf p}\varrho_k+\sigma_w^2}.
\end{align*}
The full CSI is given by 
\begin{equation*}
\hat{\boldsymbol h}_k\overset{(a)}{=}\boldsymbol{\varpi}_k\hat{\boldsymbol h}_k^{\sf p}
\overset{(b)}{=}\boldsymbol{\varpi}_k\boldsymbol h_k^{\sf p}
+\boldsymbol{\varpi}_k\boldsymbol e_k^{\sf p}
\overset{(c)}{=}\boldsymbol h_k+\boldsymbol e_k,
\end{equation*}
where $(a)$ follows from \eqref{eq:full_interpolation}, $(b)$ follows from \eqref{eq:lmmsecsi}, and $(c)$ follows by neglecting the interpolation error, i.e., $\boldsymbol h_k=\boldsymbol{\varpi}_k\boldsymbol h_k^{\sf p}$ \cite{hsieh1998channel}. 
In terms of each element, we have \eqref{eq:csi_error_model}.
Since $\rho_{k,m} =\|\boldsymbol g_{k,m}\|^2$, we have \eqref{eq:es_k} and \eqref{eq:es_e}. This completes the proof.

\bibliographystyle{IEEEtran}
\bibliography{reference/mybib}
\end{document}